\documentclass{ws-mpla}

\begin{document}

\markboth{Sini R and V C Kuriakose } {Quasinormal modes of RN black
hole space time with cosmic string in a Dirac field}

%
\catchline{}{}{}{}{}
%

\title{Quasinormal modes of RN black hole space time with cosmic string in a Dirac field}

\author{SINI R}

\address{Department of Physics, Cochin University of Science and Technology,
Kochi 682022, India.
\\
 sini@cusat.ac.in}

\author{V C KURIAKOSE}

\address{Department of Physics, Cochin University of Science and Technology,
Kochi 682022, India.
\\
 vck@cusat.ac.in}

\maketitle

\pub{Received (Day Month Year)}{Revised (Day Month Year)}

\begin{abstract}
We evaluate quasinormal mode frequencies for RN black hole space
times with cosmic string perturbed by a massless Dirac field, using
P\"{o}schl-Teller potential method. We find that only in the case of
RN black hole having small charge, the effect due to cosmic string
will dominate when perturbed by a negatively charged Dirac field,
but if we are perturbing with positively charged dirac field decay
will be less in the case of black hole having cosmic string compared
to the RN black hole without string.. 
\end{abstract}

\ccode{PACS numbers:04.70.-s, 04.62.+v , 11.27.+d}

 \maketitle
\section{Introduction}
There has been a great interest in the long standing issue in
classical relativistic theory of gravitation: small perturbations or
quasinormal modes associated with black holes. The question of
stability of black hole was first treated by Regge and Wheeler
\cite{1f} who investigated linear perturbations of the exterior
Schwarzschild space time. Further work \cite{1g} on this problem led
to the study of quasinormal modes which is believed as a
characteristic sound of black holes. Quasinormal modes (QNMs)
describe the damped oscillations under perturbations in the
surrounding geometry of a black hole with frequencies and damping
times of oscillations entirely fixed by the black hole parameters.
The study of QNMs became an intriguing subject of discussion for
last few decades \cite{nlt99,kdk99,1h,bw05} and references therein.
QNMs carry unique finger prints of black holes and it is well known
that they are crucial in studying the gravitational and
electromagnetic perturbations around black hole space times. They
are also seem to have an observational significance as the
gravitational waves produced by the perturbations, in principle, can
be used for unambiguous detection of black holes. This motivates us
to study the quasinormal mode spectra of black holes.

The motivation of the present work is to study the signature of
cosmic strings on QNMs. It has been recognized that certain gauge
theories allow the possibility of topological defects, such as
strings, magnetic poles etc and that these defects represent objects
which might have been created in the very early universe \cite{1a}.
Cosmic strings are strand of matter which could be created in a
cosmological phase transition. In 1976  Kibble suggested the
possibility of strings in the early universe\cite{1b}. These cosmic
strings might be responsible for large-scale structure in the
universe. Although little is known about these strings, it is clear
that they raise a number of issues in fundamental physics and thus
it seems to be of particular interest, both as a possible "seed" for
galaxy formation \cite{1c,1d} and as a possible gravitational
lens\cite{1e}.

The QNMs of scalar perturbations around a Schwarzschild black hole
pierced by a cosmic string was done earlier \cite{sch07}. We have
evaluated earlier\cite{sr08} quasinormal mode frequencies for
Schwarzschild, RN extremal, SdS and near extremal SdS black hole
space times with cosmic string perturbed by a massless Dirac field.
In this paper we study the influence of cosmic string on the QNMs of
RN black hole background space time which are perturbed by a
massless Dirac field. In section II, we consider the Dirac equation
in a RN black hole space time with a cosmic sting and its deduction
into a set of second order differential equations. In Section III we
evaluate the Dirac quasinormal frequencies for the massless case
using P\"{o}schl-Teller potential method for RN black holes.

\section{RN black hole with cosmic string space time for a Dirac field}
 The metric describing a charged spherically symmetric black
hole(RN) with a cosmic string can be written as\cite{ee},
\begin{equation}
ds^{2}=f(r)dt^{2}-\frac{dr^{2}}{f\left( r\right) }-r^{2}d\theta
^{2}-b^{2}r^{2}\sin ^{2}\theta d\phi ^{2}  \label{e}.
\end{equation}%
Here $f(r)=1-\frac{2M}{r}+\frac{Q^{2}}{r^{2}}$, $Q$ and $M$ are
representing the electric charge and the mass of the black hole.
This can be constructed by removing a wedge, which is done by
requiring that the azimuthal angle around the axis runs over the
range $0<\phi'<2\pi b$, with $\phi'=b\phi$ where $\phi$ runs over
zero to $2\pi$. Here $b=1-4\check{\mu }$ with $\check{\mu}$ being
the linear mass density of the string. Using the procedure adopted
in reference \cite{jnwheler} the Dirac equation in the presence of
an electromagnetic interaction in a general background space time
can be written as,
\begin{equation}
\left( \gamma ^{\mu }(\partial _{\mu }-\Gamma _{\mu }-\imath e
A_{\mu})+m\right) \Psi =0  \label{a},
\end{equation}
where $m$ and $e$ are the the mass and charge of the Dirac field,
$A_{\mu}$ is the electro magnetic potential which can be written as,
\begin{equation}
A_{\mu}=\left(\frac{Q}{r},0,0,0\right)\label{177},
\end{equation}
Here
\begin{equation}
\gamma^{\mu}=g^{\mu\nu}\gamma_{\nu},
\end{equation}
and
\begin{equation}
\gamma_{\nu}=e_{\nu}^{a}\gamma_{a},
\end{equation}
where $\gamma^{a}$ are the Dirac matrices,
\begin{equation}
\gamma_{0}=\left[\begin{array}{cc}
\imath  &0 \\
0 & -\imath%
\end{array}%
\right],\gamma_{i}=\left[\begin{array}{cc}
0 &\sigma_{i} \\
\sigma_{i} &0%
\end{array}%
\right],
\end{equation}
and $\sigma_{i}$ are the Pauli matrices. $ e_{\nu}^{a}$ is the
tetrad given by,
\begin{equation}
e_{t}^{t}=f^{\frac{1}{2}};e_{r}^{r}=\frac{1}{f^{\frac{1}{2}}};e_{\theta
}^{\theta }=r,e_{\phi }^{\phi }=br\sin \theta  \label{111}.
\end{equation}
The inverse of the tetrad $ e_{\nu}^{a }$ is defined by,
\begin{equation}
g^{\mu\nu}=\eta^{ab}e_{a}^{\mu }e_{b}^{\nu },
\end{equation}
with $\eta^{ab}=diag(-1,1,1,1)$, the Minkowski metric. The spin
connection $\Gamma_{\mu} $ is given by
\begin{equation}
\Gamma _{\mu}=-\frac{1}{2}[\gamma_{a},\gamma_{b}] e_{\nu
}^{a}e^{b\nu}_{ ;\mu} \label{b},
\end{equation}
where $ e^{b\nu}_{ ;\mu} =\partial _{\mu }e^{b\nu }+\Gamma _{\kappa
\mu }^{\nu }e^{b\kappa } $ is the covariant derivative of $e^{b\nu
}$. The spin connections for the above metric are obtained as,
\begin{eqnarray}
\Gamma _{t} &=&\frac{1}{4}\gamma _{1}\gamma _{0}\frac{\partial
f\left( r\right) }{\partial r},
\label{113} \\
\Gamma _{r} &=&0, \\
\Gamma _{\theta } &=&\frac{1}{2}\gamma _{1}\gamma
_{2}f^{\frac{1}{2}}\left(
r\right) , \\
\Gamma _{\phi } &=&\frac{1}{2}\gamma _{2}\gamma _{3}b\cos \theta +\frac{1}{2}%
\gamma_{1}\gamma _{3}b\sin \theta f^{\frac{1}{2}}\left( r\right).
\end{eqnarray}
Substituting the spin connections in Eq. (\ref{a}) we will get ,
\begin{equation}
[\frac{-\gamma _{0}}{f^{\frac{1}{2}}}(\frac{\partial }{\partial
t}-\imath \frac{eQ}{r})
+\gamma _{1}f^{\frac{1}{2}}( \frac{\partial }{\partial r}+\frac{1}{r}+%
\frac{1}{4{f( r) }}{\frac{\partial f}{\partial r} }) +
\frac{\gamma_{2}}{r}(\frac{\partial }{\partial \theta }+\frac{1}{2}\cot\theta)+\frac{\gamma_{3}}{%
br\sin \theta }\frac{\partial }{\partial \phi }+m]
\Psi(t,r,\theta,\phi) =0 \label{114a},
\end{equation}
Using the transformation $\Psi(t,r,\theta,\phi)=\frac{\exp(-\imath
Et)}{rf^{\frac{1}{4}}{\sin\theta}^{\frac{1}{2}}}\chi(r,\theta,\phi)$,
Eq.(\ref{114a}) becomes,
\begin{eqnarray}
[\frac{-E-\frac{e Q}{r}}{f^{\frac{1}{2}}}\frac{\partial }{\partial t}%
+\frac{1}{\imath}\gamma_{0}\gamma _{1}f^{\frac{1}{2}}(
\frac{\partial }{\partial r})  +
\frac{\gamma_{1}}{\imath r}\gamma_{1}\gamma_{0}(\gamma _{2}\frac{\partial }{\partial \theta }&+&\nonumber\\ \frac{\gamma_{3}}{%
b\sin \theta }\frac{\partial }{\partial \phi })-\imath \gamma_{0}m]
\chi(r,\theta,\phi) =0 \label{114}.
\end{eqnarray}
Dirac equation can be separated out into radial and angular parts by
the following substitution,
\begin{equation}
\chi(r,\theta,\phi)=R(r)\Omega(\theta,\phi)\label{215}.
\end{equation}
The angular momentum operator is introduced as, \cite{jnwheler}
\begin{equation}
\mathbf{K}_{\left( b\right) }=-\imath\gamma _{1}\gamma_{0}\left(
\gamma _{2}\partial _{\theta }+\gamma _{3}\left( b\sin \theta
\right) ^{-1}\partial _{\phi }\right)   \label{116},
\end{equation}
such that,
\begin{equation}
\mathbf{K}_{ (b)}\Omega(\theta,\phi)=k_{b}\Omega(\theta,\phi),
\label{a117}
\end{equation}
where the eigenvalues of $\mathbf{K}_{(b)}$ are $k_{b}=\frac{k}{b}$.
Here k is a positive or a negative nonzero integer with
$l=|k+\frac{1}{2}|-\frac{1}{2}$, where $l$ is the total orbital
angular momentum. The cosmic string presence is codified in the
eigenvalues of the angular momentum operator\cite{1j}. Substituting
Eqs.(\ref{215}) and (\ref{a117}) in Eq.(\ref{114}), we will get
radial equation which contains $\gamma _{0}$ and $\gamma _{1}$. As
$\gamma _{0}$ and $\gamma _{1}$ can be represented by $2\times2$
matrices, we write the radial factor $R(r)$ by a two component
spinor notation,
\begin{equation}
R(r)=\left[\begin{array}{c}
F \\
G%
\end{array}\right].%
\end{equation}
Then the radial equation in F an G are given by,
\begin{equation}
f\frac{dG}{dr}+f^{\frac{1}{2}}\frac{k_{b}}{r}G+f^{\frac{1}{2}}mF=(E+\frac{eQ}{r})
F \label{i},
\end{equation}%
\begin{equation}
f\frac{dF}{dr}-f^{\frac{1}{2}}\frac{k_{b}}{r}F+f^{\frac{1}{2}}mG=-(E+\frac{eQ}{r})
G  \label{j}.
\end{equation}
Let us have a co-ordinate change given by,
\begin{equation}
dr_{\ast }=\frac{dr}{f}  \label{117}.
\end{equation}%
 Eq.(\ref{i}) and Eq.(\ref{j})  then become,
\begin{eqnarray}
\frac{\partial }{\partial r_{\ast }}\left[
\begin{array}{c}
G \\
F%
\end{array}%
\right] +f^{\frac{1}{2}}\left[
\begin{array}{cc}
\frac{k_{b}}{r} & m \\
m & -\frac{k_{b}}{r}%
\end{array}%
\right] \left[
\begin{array}{c}
G \\
F%
\end{array}%
\right] =\left[
\begin{array}{cc}
0 & E+\frac{e Q}{r} \\
-E-\frac{e Q}{r}  & 0%
\end{array}%
\right] \left[
\begin{array}{c}
G \\
F%
\end{array}%
\right]    \label{118}.
\end{eqnarray}
Defining,
\begin{equation}
\left[
\begin{array}{c}
\hat{G} \\
\hat{F}
\end{array}%
\right] =\left[
\begin{array}{cc}
\cos \frac{\theta}{2} & -\sin \frac{\theta}{2} \\
\sin \frac{\theta}{2} & \cos \frac{\theta}{2}%
\end{array}%
\right] \left[
\begin{array}{c}
G \\
F
\end{array}%
\right]  \label{119},
\end{equation}
where
\begin{equation}
\theta =\tan ^{-1}(\frac{mr}{|k_{b}|})  \label{120}.
\end{equation}%
Eq.(\ref{118}) becomes,
\begin{eqnarray}
\frac{\partial }{\partial r_{\ast }}\left[
\begin{array}{c}
\hat{G} \\
\hat{F}%
\end{array}%
\right]
+f^{\frac{1}{2}}\sqrt{\left(\frac{k_{b}}{r}\right)^{2}+m^{2}}\left[
\begin{array}{cc}
1 & 0 \\
0 & -1%
\end{array}%
\right] \left[
\begin{array}{c}
\hat{G} \\
\hat{F}%
\end{array}%
\right]\nonumber\\ =-E \left[ 1+\frac{1}{2E }\frac{fm|k_{b}|+}{{%
{k_{b}}^{2}+m^{2}r^{2}}}++\frac{e Q}{r}\right] \left[
\begin{array}{cc}
0 & -1 \\
1 & 0%
\end{array}%
\right] \left[
\begin{array}{c}
\hat{G} \\
\hat{F}%
\end{array}%
\right]   \label{121}.
\end{eqnarray}
By making another change of the variable;
\begin{equation}
d\hat{r}_{\ast }=\frac{dr_{\ast }}{\left[ 1+\frac{1}{2E }\frac{fmk_{b}}{{%
k_{b}^{2}+m^{2}r^{2}}}+e\frac{Q}{E r}\right] }  \label{122}.
\end{equation}
Eq.(\ref{121}) can be simplified to,
\begin{equation}
\frac{\partial }{\partial \hat{r}_{\ast }}\left[
\begin{array}{c}
\hat{G} \\
\hat{F}%
\end{array}%
\right] +\frac{f^{\frac{1}{2}}\sqrt{\left(\frac{k_{b}}{r}\right)^{2}+m^{2}}}{\left[ 1+\frac{1}{%
2E }\frac{fm|k_{b}|}{{k_{b}^{2}+m^{2}r^{2}}}+e\frac{Q}{E r}\right]
}\left[
\begin{array}{cc}
1 & 0 \\
0 & -1%
\end{array}%
\right] \left[
\begin{array}{c}
\hat{G}\\
\hat{F}%
\end{array}%
\right] =E \left[
\begin{array}{cc}
0 & 1 \\
-1 & 0%
\end{array}%
\right] \left[
\begin{array}{c}
\hat{G} \\
\hat{F}%
\end{array}%
\right]   \label{123},
\end{equation}
i.e,
\begin{equation}
\frac{\partial }{\partial \hat{r}_{\ast }}\left[
\begin{array}{c}
\hat{G} \\
\hat{F}%
\end{array}%
\right] +W\left[
\begin{array}{c}
\hat{G} \\
-\hat{F}%
\end{array}%
\right] =E \left[
\begin{array}{c}
\hat{F} \\
-\hat{G}%
\end{array}%
\right]   \label{124},
\end{equation}
where
\begin{equation}
W=\frac{f^{\frac{1}{2}}\sqrt{\left(\frac{k_{b}}{r}\right)^{2}+m^{2}}}{\left[ 1+e\frac{Q}{E r}+\frac{1}{2E }%
\frac{fmk_{b}}{{k_{b}^{2}+m^{2}r^{2}}}\right] }  \label{125}.
\end{equation}%
Thus from Eq.(\ref{124}), we will get coupled equations for
$\hat{G}$ and $\hat{F}$ which are given bellow,
\begin{equation}
-\frac{\partial ^{2}\hat{F}}{\partial \hat{r}_{\ast
}^{2}}+V_{1}\hat{F}=E ^{2}\hat{F} \label{126},
\end{equation}
\begin{equation}
-\frac{\partial ^{2}\hat{G}}{\partial \hat{r}_{\ast
}^{2}}+V_{2}\hat{G}=E^{2}\hat{G} \label{127},
\end{equation}
where
\begin{equation}
V_{1,2}=\pm \frac{\partial W}{\partial \hat{r}_{\ast }}+W^{2}
\label{128}.
\end{equation}
From the equations Eqs (\ref{126}) and (\ref{127}), we can evaluate
the corresponding quasinormal mode frequencies for various black
hole space time. Here $V_{1}$ and $V_{2}$ are the super symmetric
partners derived from the same super potential $W$ \cite{aa}. It has
been shown that potentials related in this way possess the same
spectra of quasi-normal mode frequencies.
\section{Quasinormal mode frequencies}

 We shall evaluate the quasi-normal frequencies for RN black hole space times perturbed
by a massless Dirac field. For massless case, i.e, $m=0$, the
equation for the potential in Eq.(\ref{128}) becomes,

\begin{eqnarray}
V= f\left(1+e\frac{Q}{E r}\right)\frac{\partial}{\partial r} \left( f^{\frac{1}{2}}\frac{k_{b}}{r\left(1+e\frac{Q}{E r}\right)}\right) %
\nonumber\\+f\left(\frac{k_{b}}{r}\right)^{2}
\frac{1}{\left(1+e\frac{Q}{E r}\right)^{2}} \label{134},
\end{eqnarray}
where we have avoided the subscript of $V_{1}$. Here we take $Q$ as
always positive and the sign of $e$ will change to positive or
negative so that the positive sign of $e$ actually means that the
product $eQ>0$ and vice versa.

\begin{figure}[h]
\center
\includegraphics[width=5.5cm]{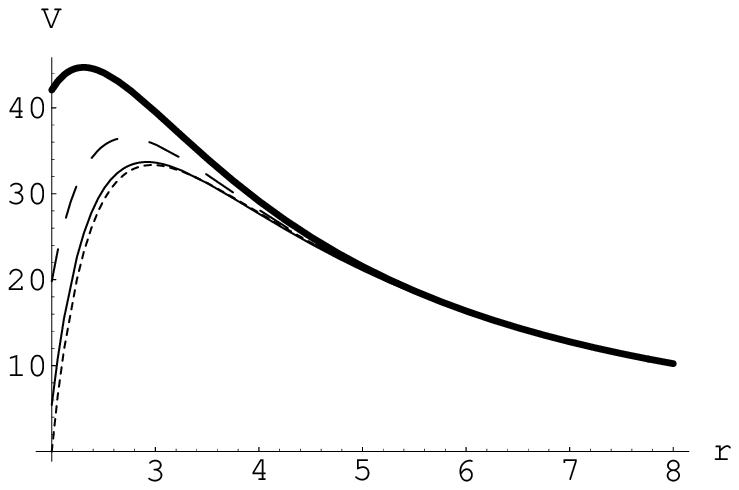}
\includegraphics[width=5.5cm]{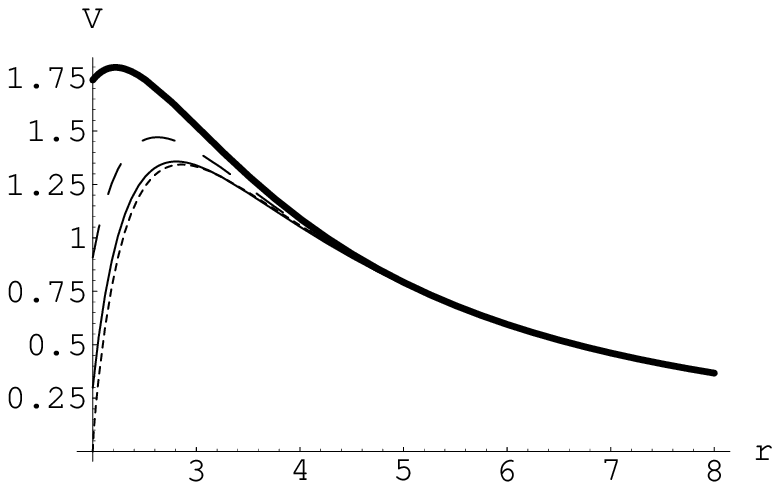}
\includegraphics[width=5.5cm]{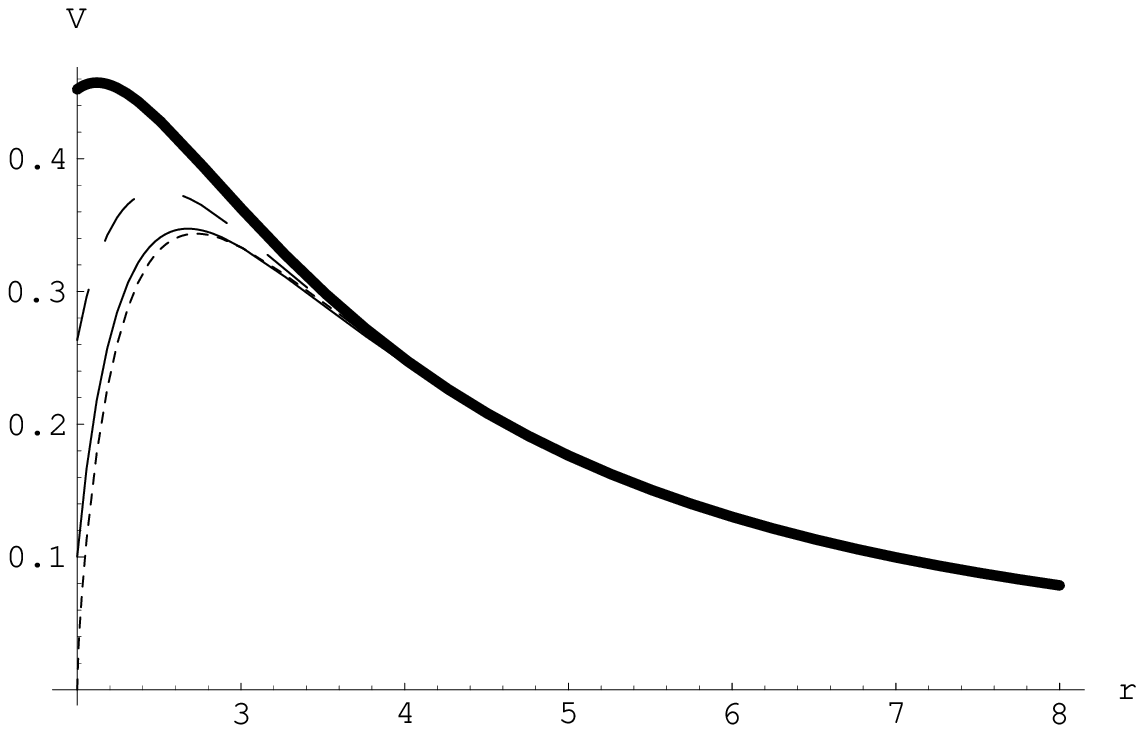}
\caption{Variation of effective potential for Dirac field with m=0
k=3, e=0.1, E=1 and Q=0(doted line), 0.3(solid line), 0.6(dashed
line)and 0.9(bold line). for each b value(0.1, 0.5 and
1)}\label{graph8}
\end{figure}
The effective potential as a function of r is plotted in
Fig.\ref{graph8}. We can see that the dependence of V on $Q$ is
strong and the peak of the potential increases faster and faster
with Q. When the value of $b$ is changed form $b=1$ to $b=0.5$ and
to $b=0.1$ the peak of the potential is increased, i.e, when the
cosmic string is present the height of the potential barrier is
increased.

 Now we shall evaluate the QNMs using P\"{o}schl-Teller potential
approximation proposed by  Ferrari and Mashhoon, the
P\"{o}schl-Teller potential \cite{vf84} is given by,
 \begin{equation}
\label{135}V_{PT}=\frac{V_{0}}{\cosh^{2}\left(\frac{r^\star}{b}\right)}.
 \end{equation}
The quantity $V_{0}$ and $b$ are given by the height and curvature
of the potential at its maximum($r=r_{max}$). Thus
\begin{equation}
\label{136}V_{0}=V_{r_{max}},
\frac{1}{b^{2}}=-\frac{1}{2V_{0}}\left[\frac{d^{2}V}{dr^{2}}\right]_{r=r_{max}}.
 \end{equation}
The QNMs of the P\"{o}schl-Teller potential can be evaluated
analytically;
\begin{equation}
\label{137}E=\frac{1}{b}\left[\sqrt{V_{0}-\frac{1}{4}}-\left(n+\frac{1}{2}\right)\imath
\right].
\end{equation}
In this case the effective potential depends both on $Q$ and $E$. So
we calculate the Quasinormal modes. Here we first find QNMs for the
case $Q=0$ in which the potential is independent of $E$, let it be
$E_{0}$. We take this as the initial value $E_{0}$ for a fixed $n$,
$l$(or $k$) and $e$ and is used to evaluate the corresponding QNMs
for $Q\neq0$. i.e, we use $E_{0}$ as real to modify the potential
and find $E_{1}$ and repeat the process successfully to get $E_{2}$,
$E_{3}$, $E_{4}$....\cite{yuwu04}.

\begin{figure}[h]
\center
\includegraphics[width=5.5cm]{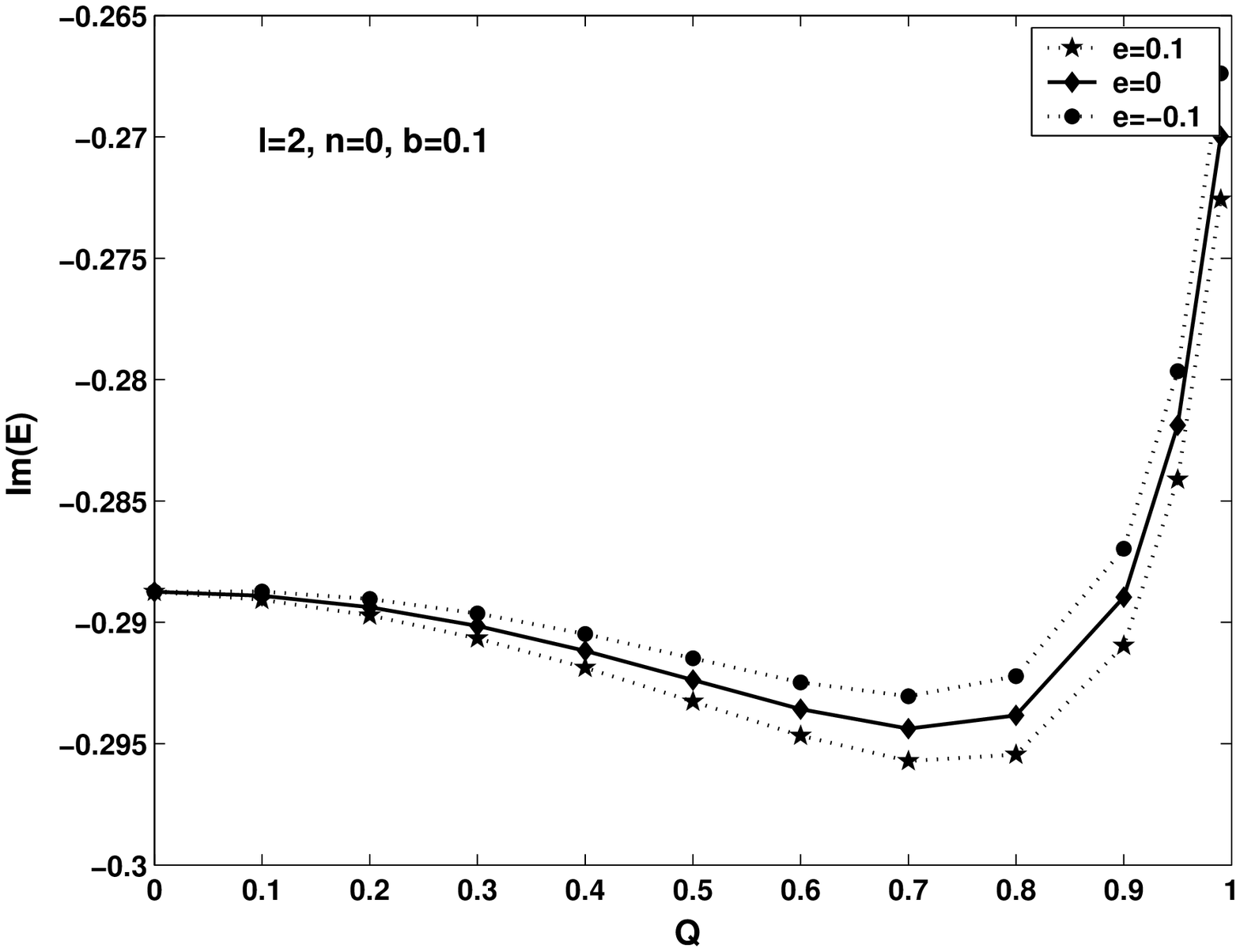}
\includegraphics[width=5.5cm]{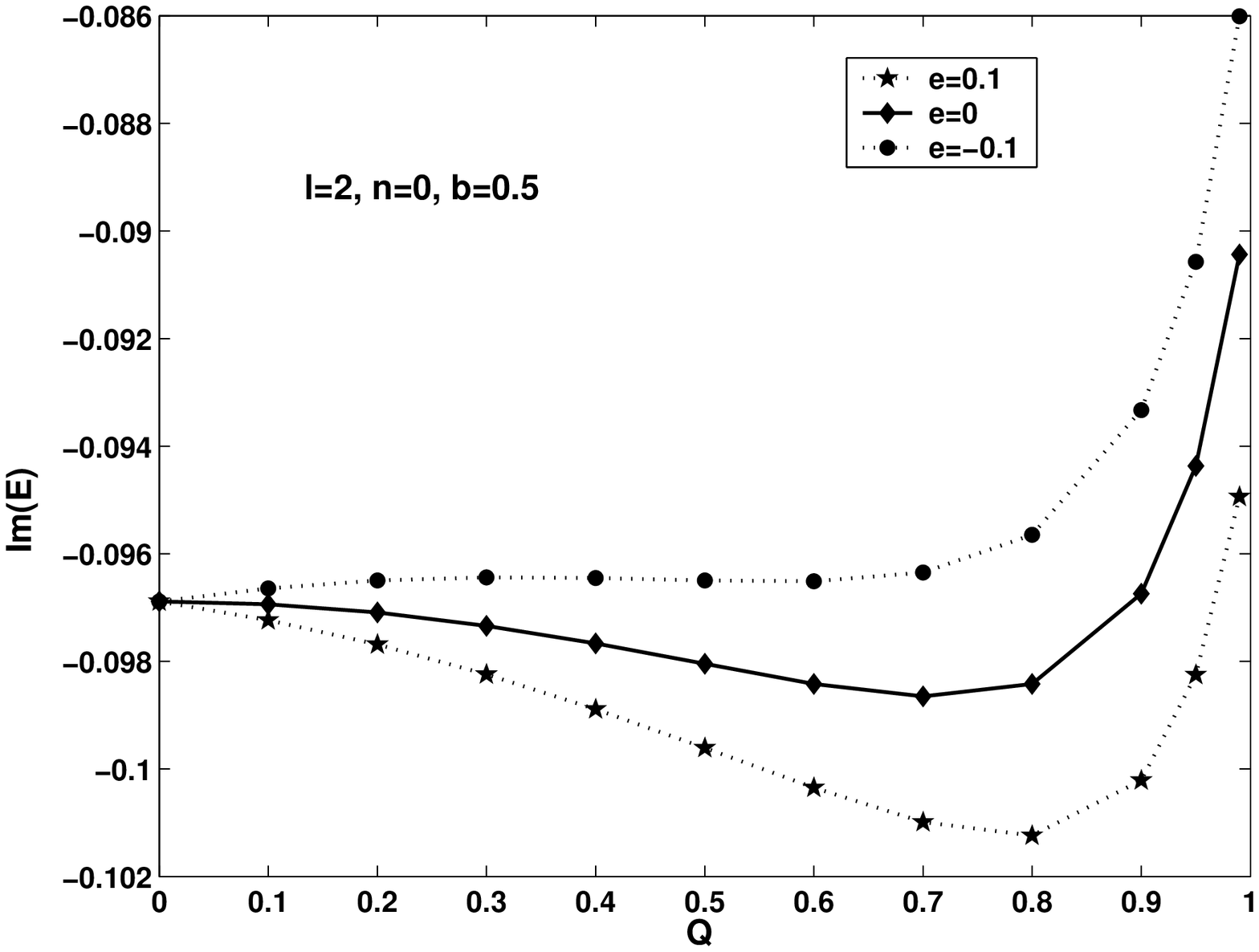}
\includegraphics[width=5.5cm]{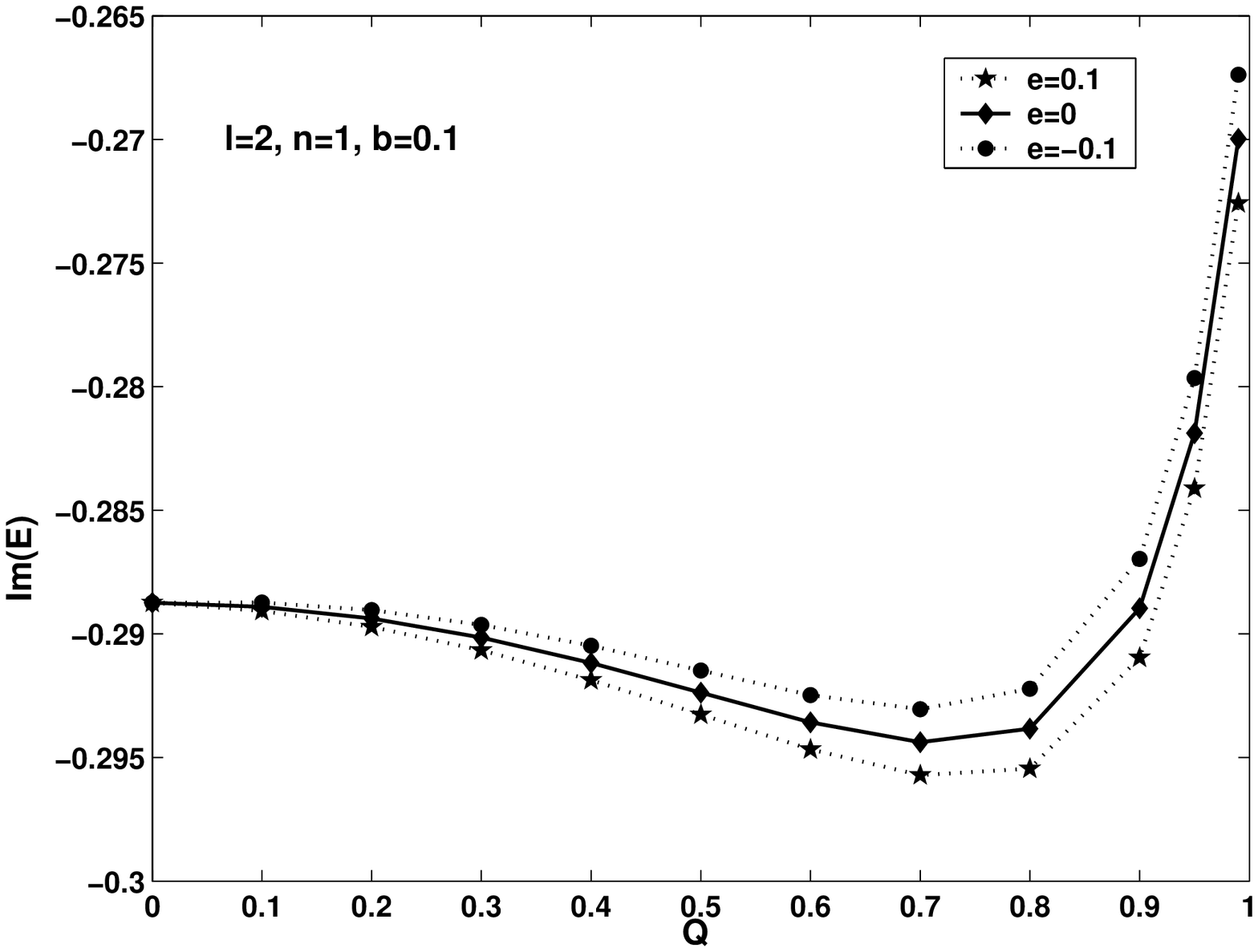}
\includegraphics[width=5.5cm]{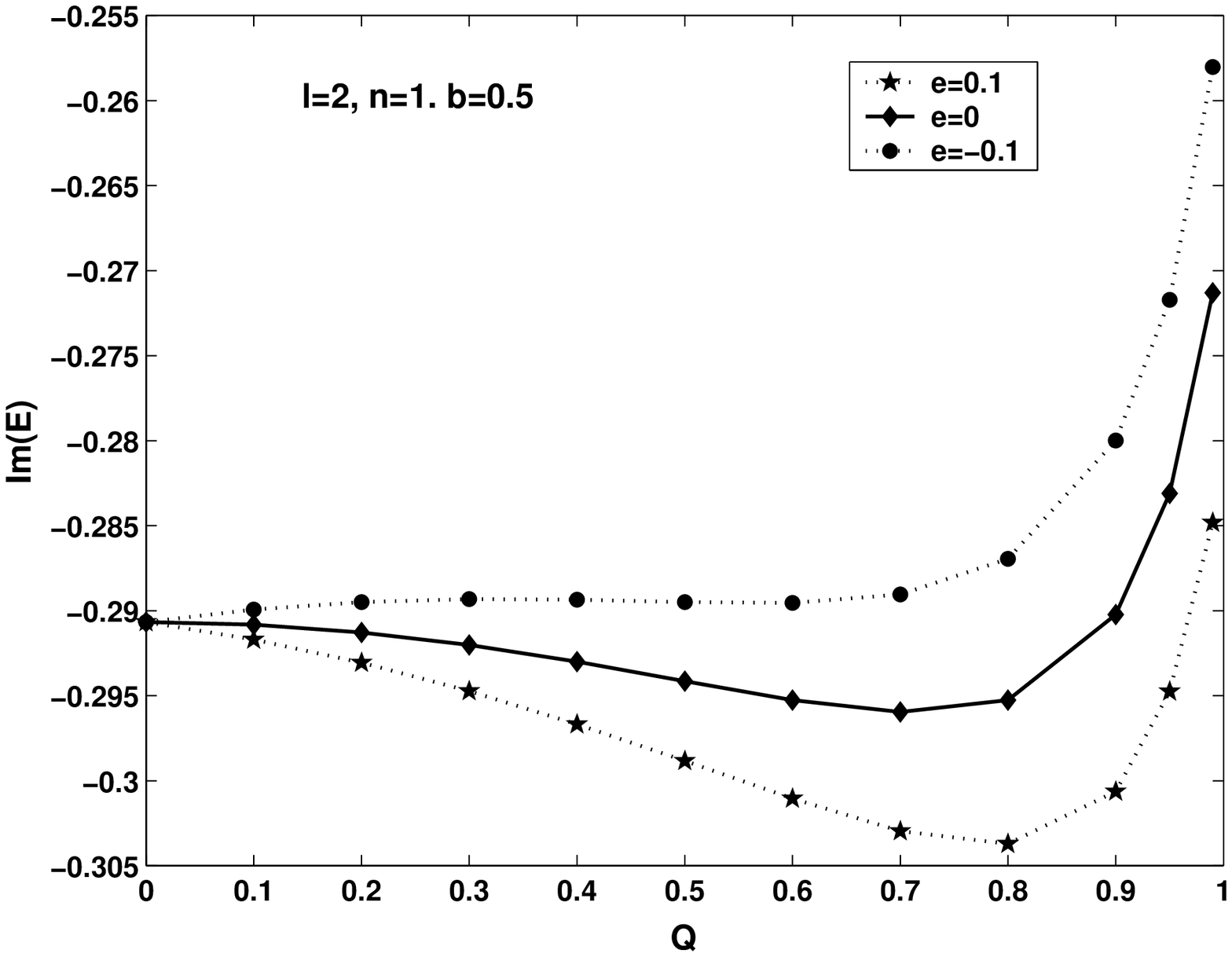}
\caption{Im(E) vs Q for b=0.1 and 0.5 for l=2, n=0 and
n=1}\label{graph2}
\end{figure}

We first checked the effect of positive and negative Dirac field
charges $'e'$ on a positively charged RN black hole for a fixed $b$
value. For that we plot Im(E) with black hole charge $Q$ for
different values of field charge $e$ for each $b$ values.
\begin{figure}[h]
\center
\includegraphics[width=5.5cm]{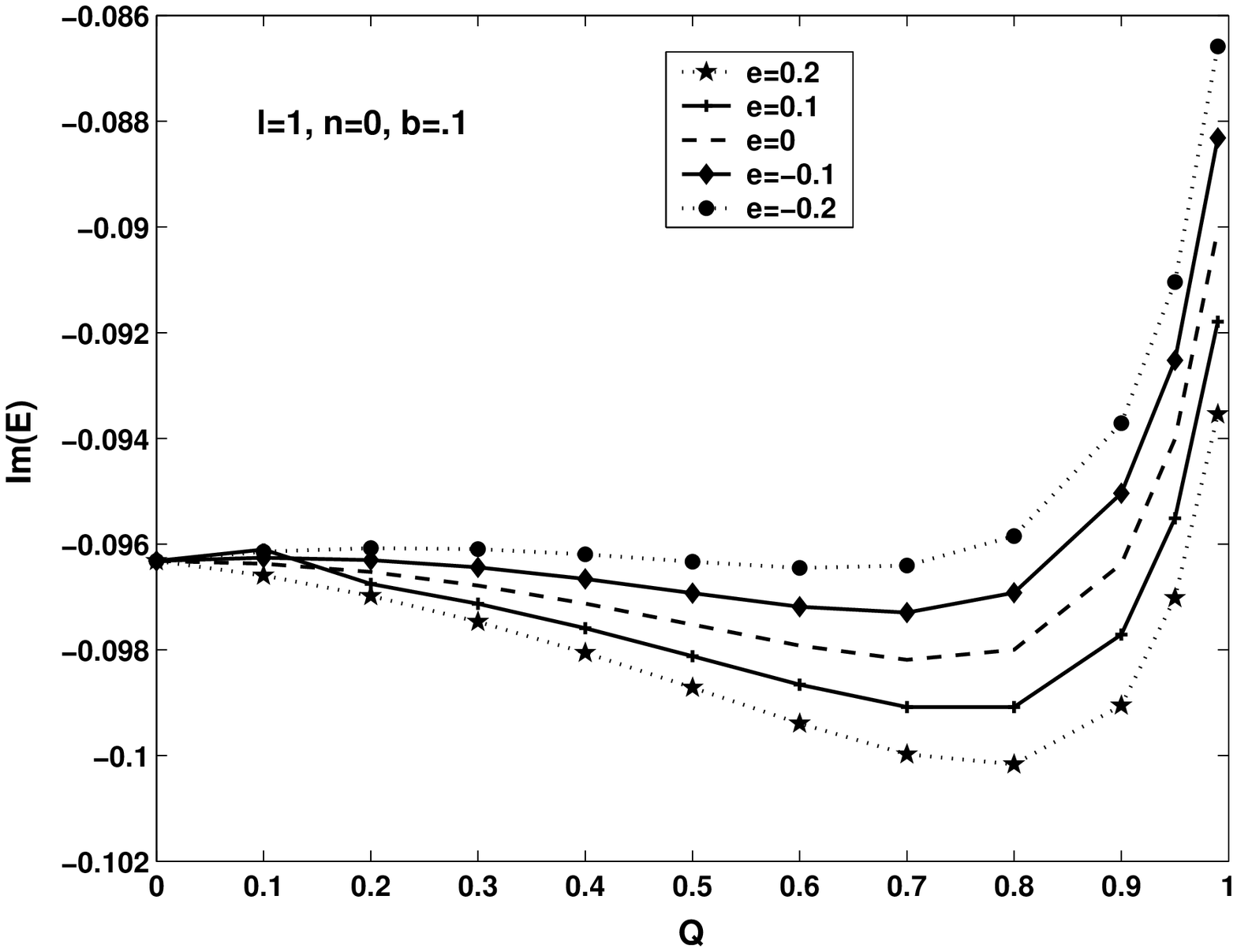}
\includegraphics[width=5.5cm]{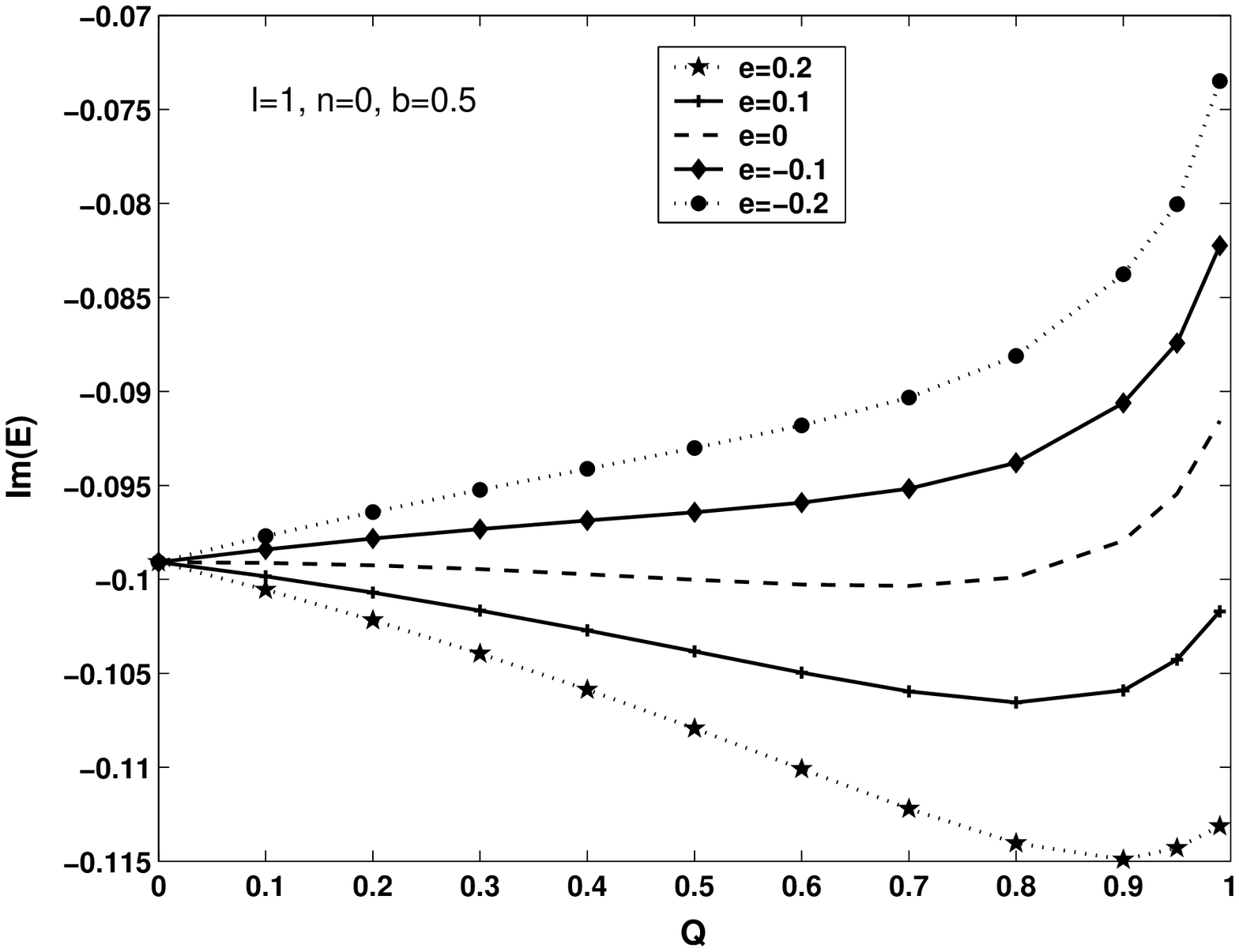}
\includegraphics[width=5.5cm]{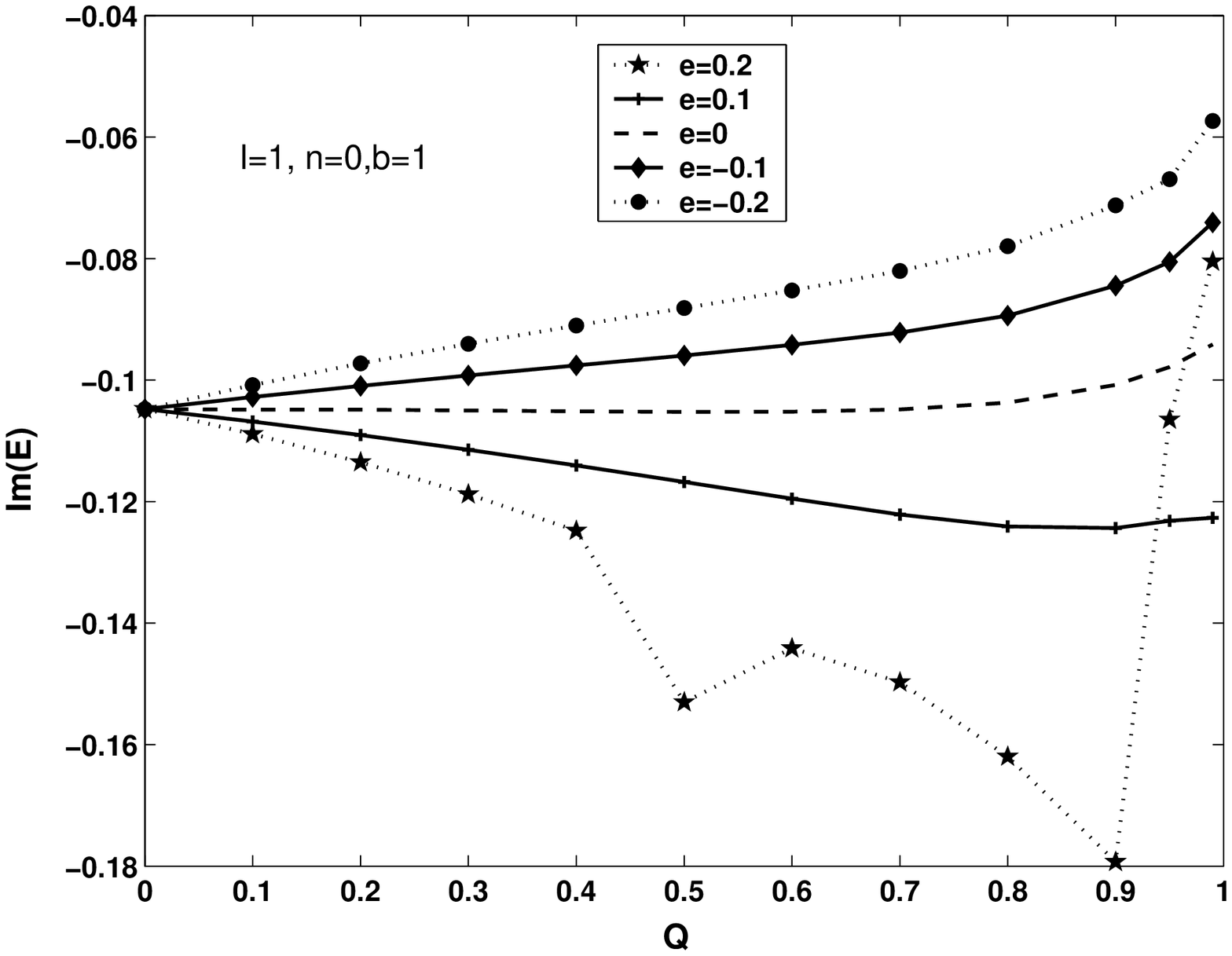}
\caption{Im(E) vs Q for l=1 and n=0 for each b value(0.1, 0.5 and
1)}\label{graph1}
\end{figure}
Fig.\ref{graph1} shows the behavior of $Im(E)$ versus Q for the mode
$l=1$($k=1$) and $n=0$ case.  And it is clear that as black hole
charge $Q$ increases from 0 to 0.99, at first there is a small
increase of $|Im(E)|$ and then it decreases with increase of the
charge $Q$. Also for a fixed $b$ value the $|Im(E)|$ is small for
negatively charged Dirac field compared to positively charged ones.
In $b=0.1$ case that is when the effect of cosmic string is high we
get very good curves compared to others. When the $l$ value
increases the behavior is similar to the former case. Fig.
\ref{graph2} shows the behavior of $Im(E)$ versus $Q$ for $l=2$
case. Here we considered both the modes $n=1$ and $n=0$ for $b=0.1$
and $b=0.5$. Thus from this we understand that positively charged
Dirac field decay faster than negatively charged Dirac field on a
positively charged RN black hole with a fixed $b$ value.

\begin{figure}[h]\center
\includegraphics[width=5.5cm]{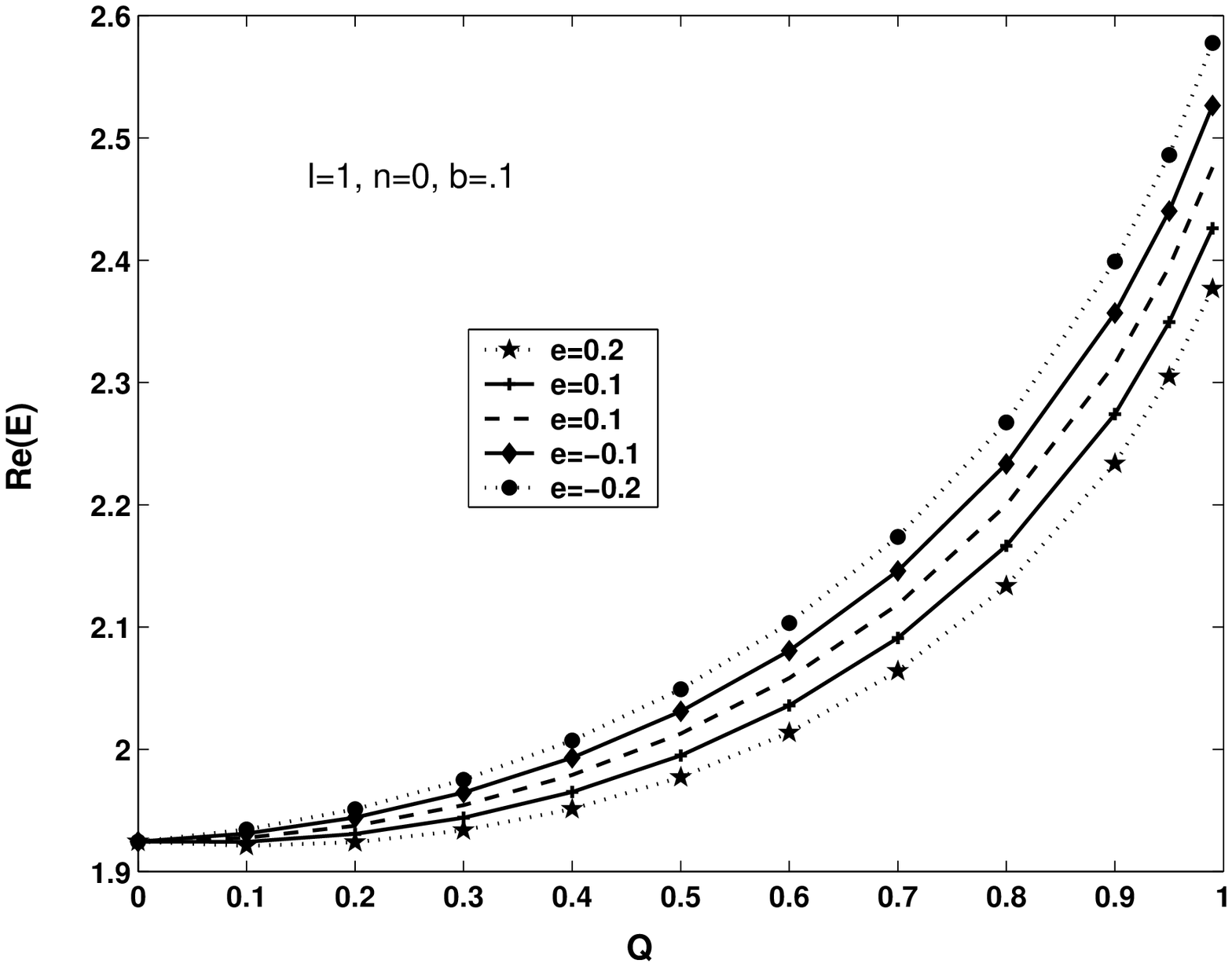}
\includegraphics[width=5.5cm]{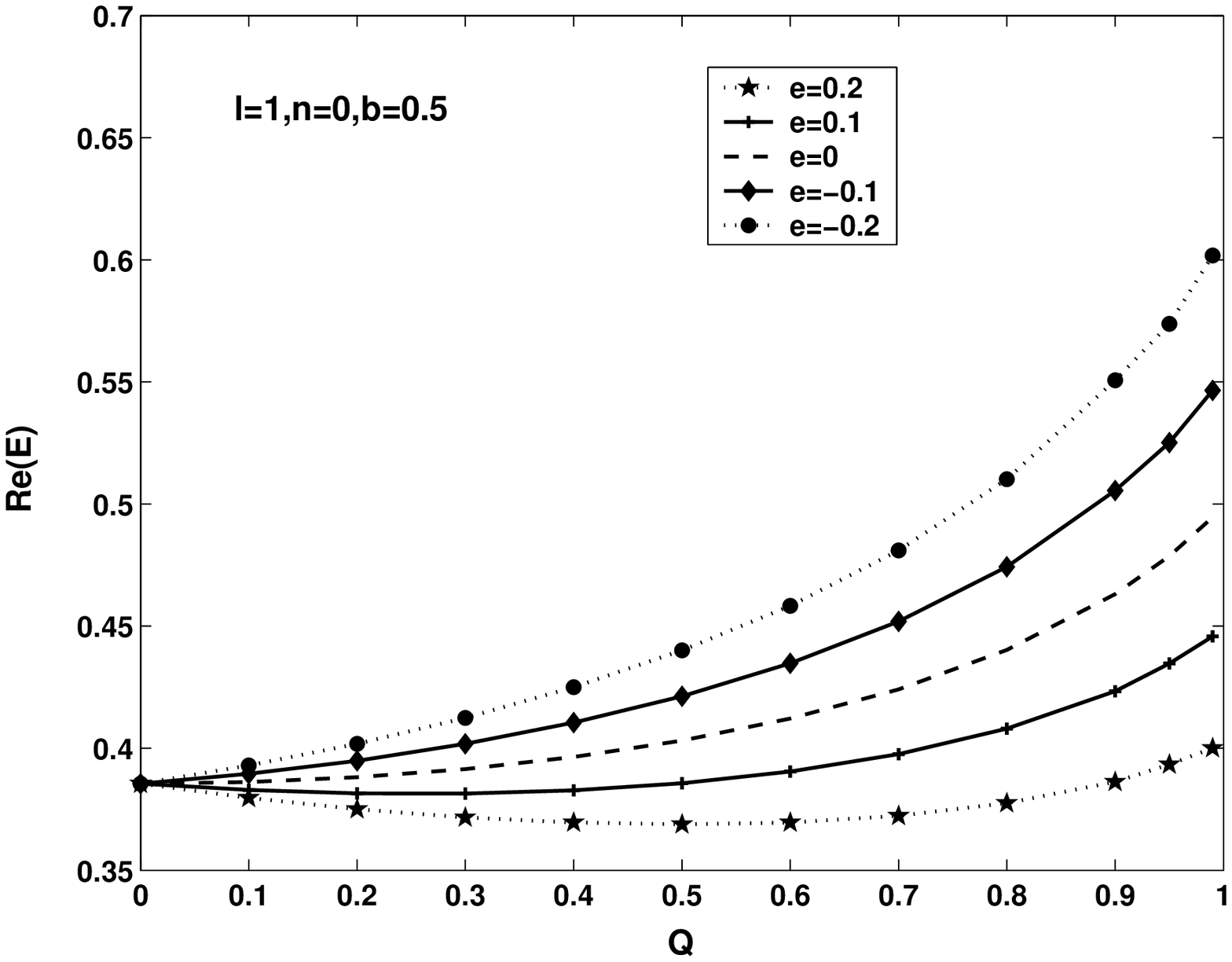}
\includegraphics[width=5.5cm]{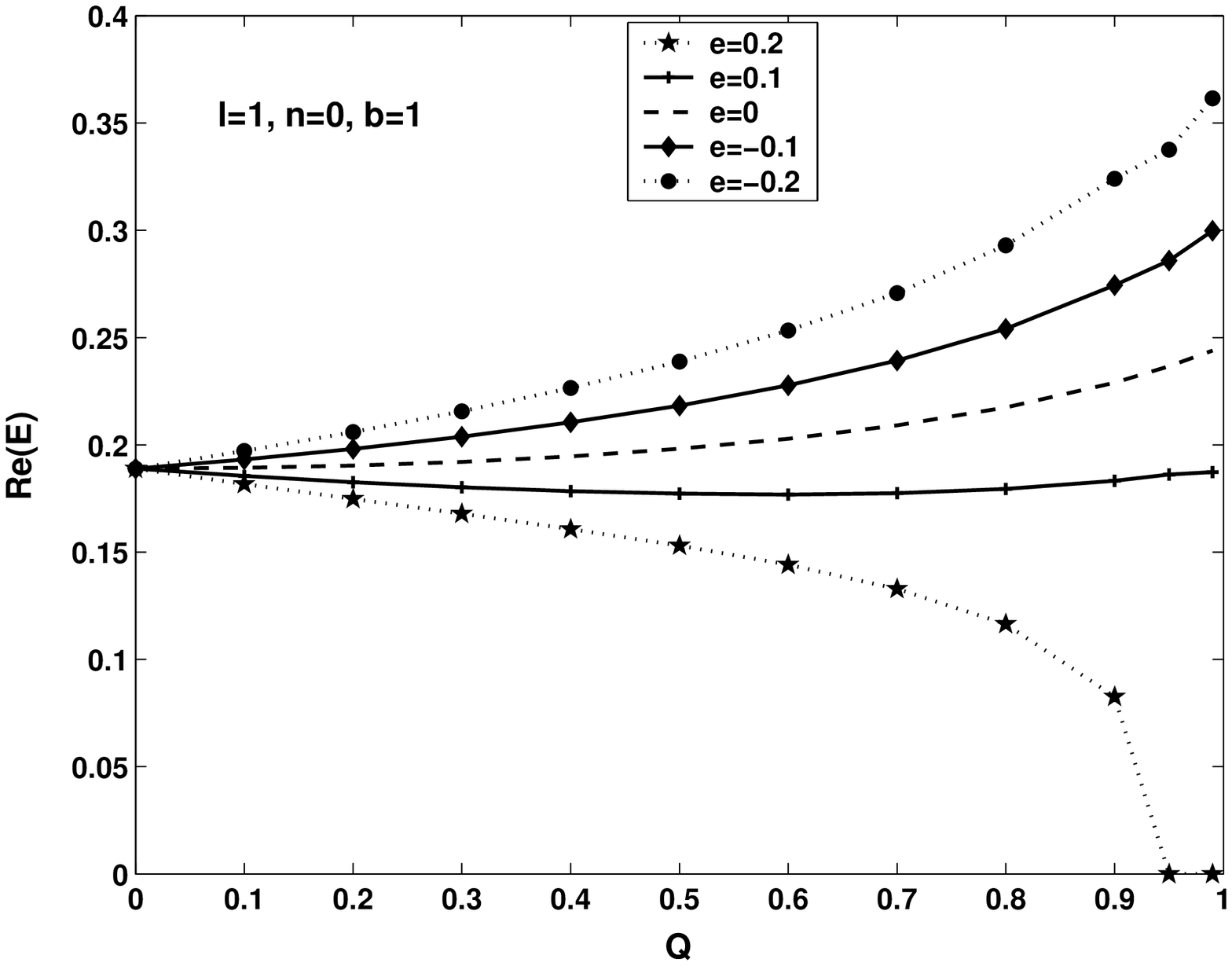}
\caption{Re(E) vs Q for l=1 and n=0 for fixed values of b(0.1, 0.5,
1)}\label{graph3}
\end{figure}

Fig. \ref{graph3} shows the dependence of $Re(E)$ with black hole
charge $Q$ with different values of Dirac field charge $e$ from +0.2
to -0.2 for fixed values of $b$. For all $b$ values $Re(E)$ is
increasing with respect to $Q$. From the Fig.\ref{graph3} it is
clear that $e=-0.2$ have higher $Re(E)$ compared to $e=0.2$ which
implies that negatively charged Dirac field have larger $Re(E)$
value.

\begin{figure}[h]\center
\includegraphics[width=5.5cm]{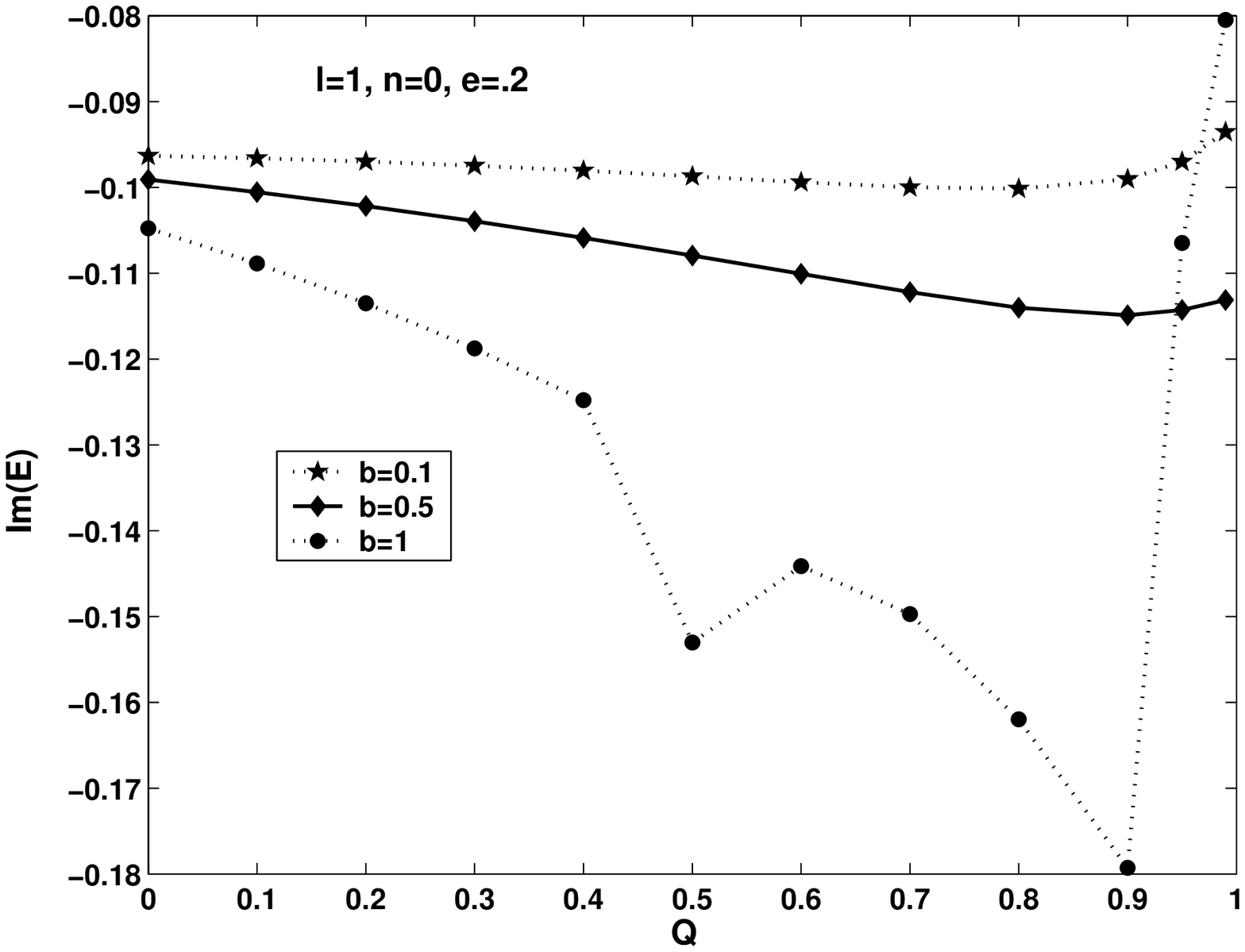}
\includegraphics[width=5.5cm]{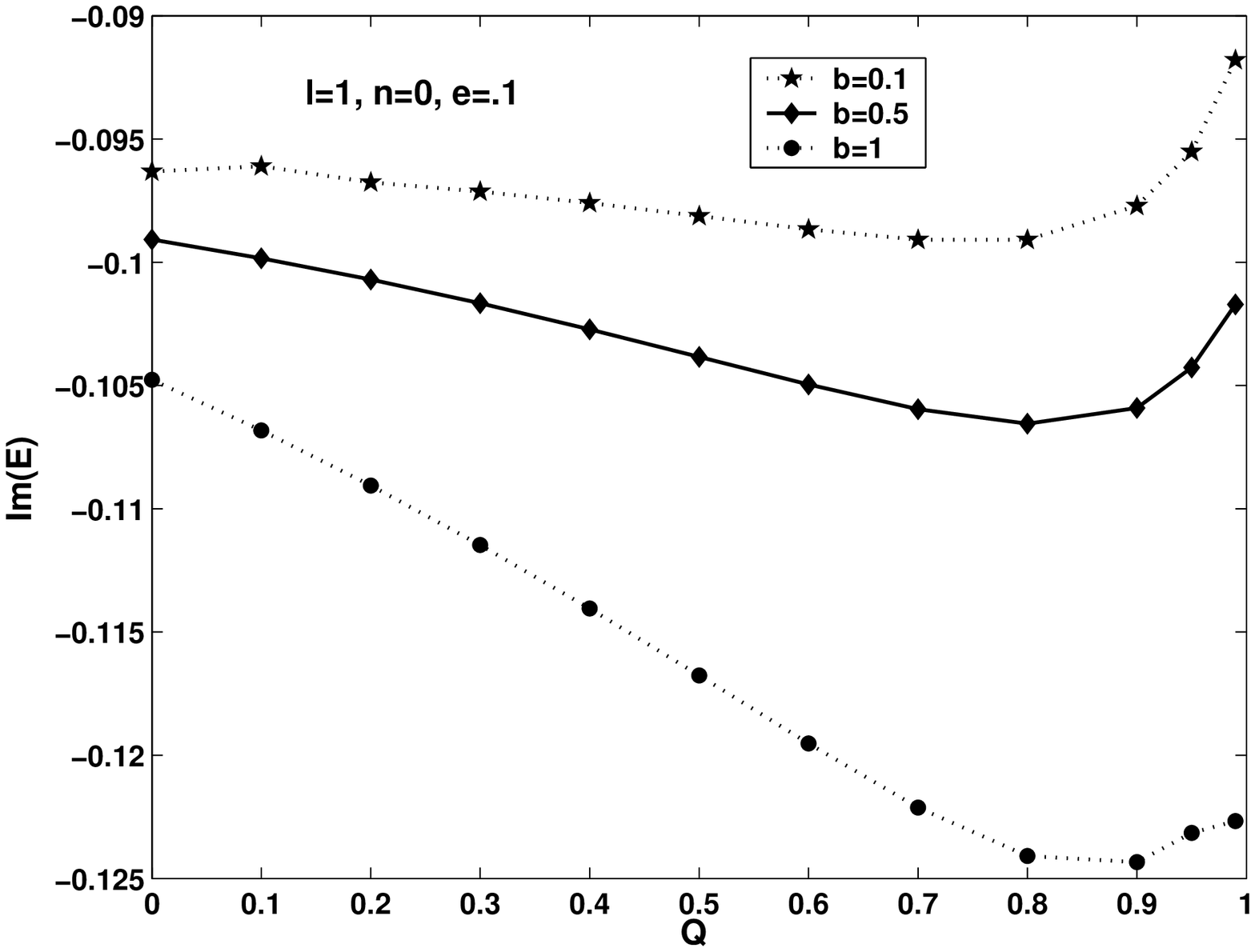}
\includegraphics[width=5.5cm]{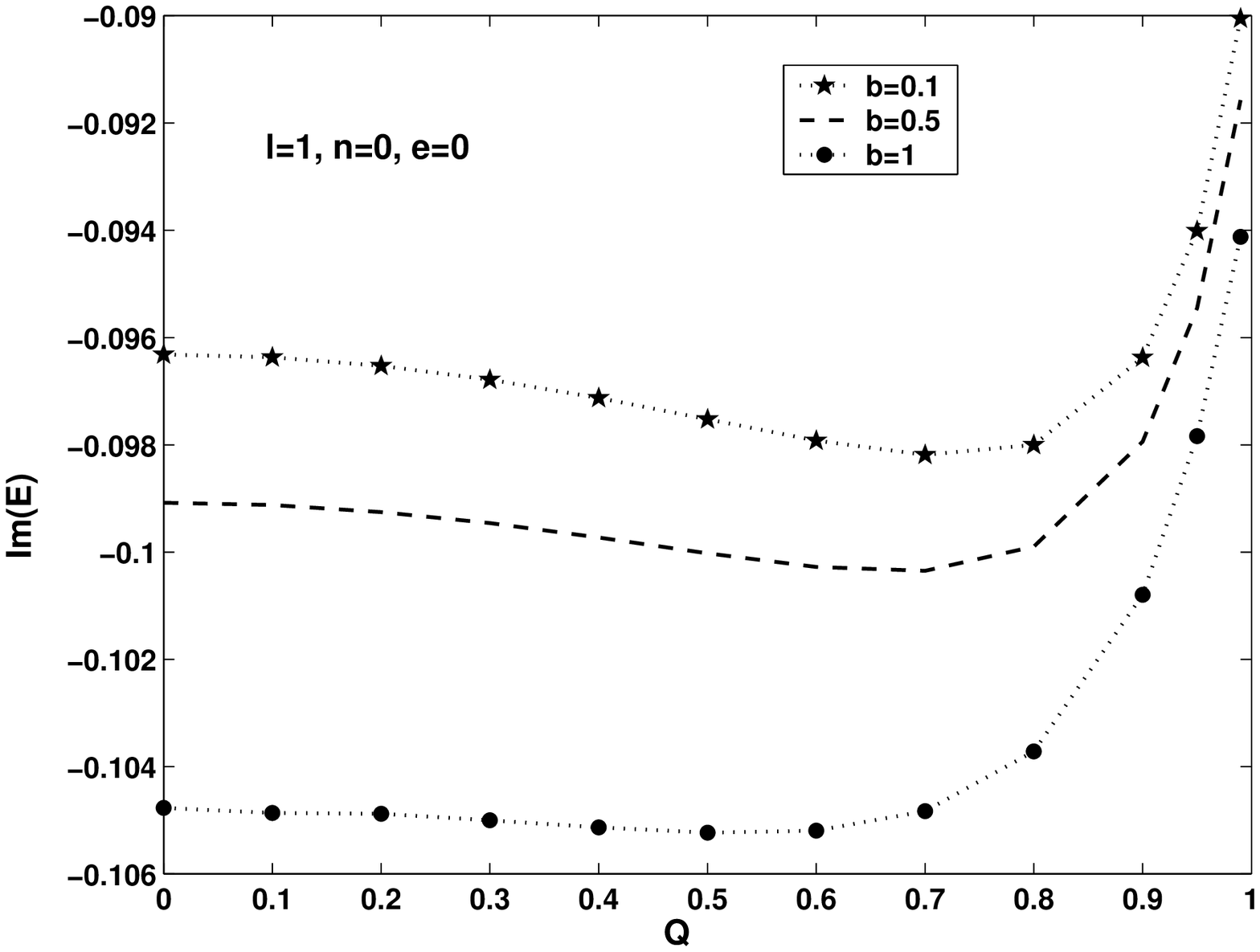}
\includegraphics[width=5.5cm]{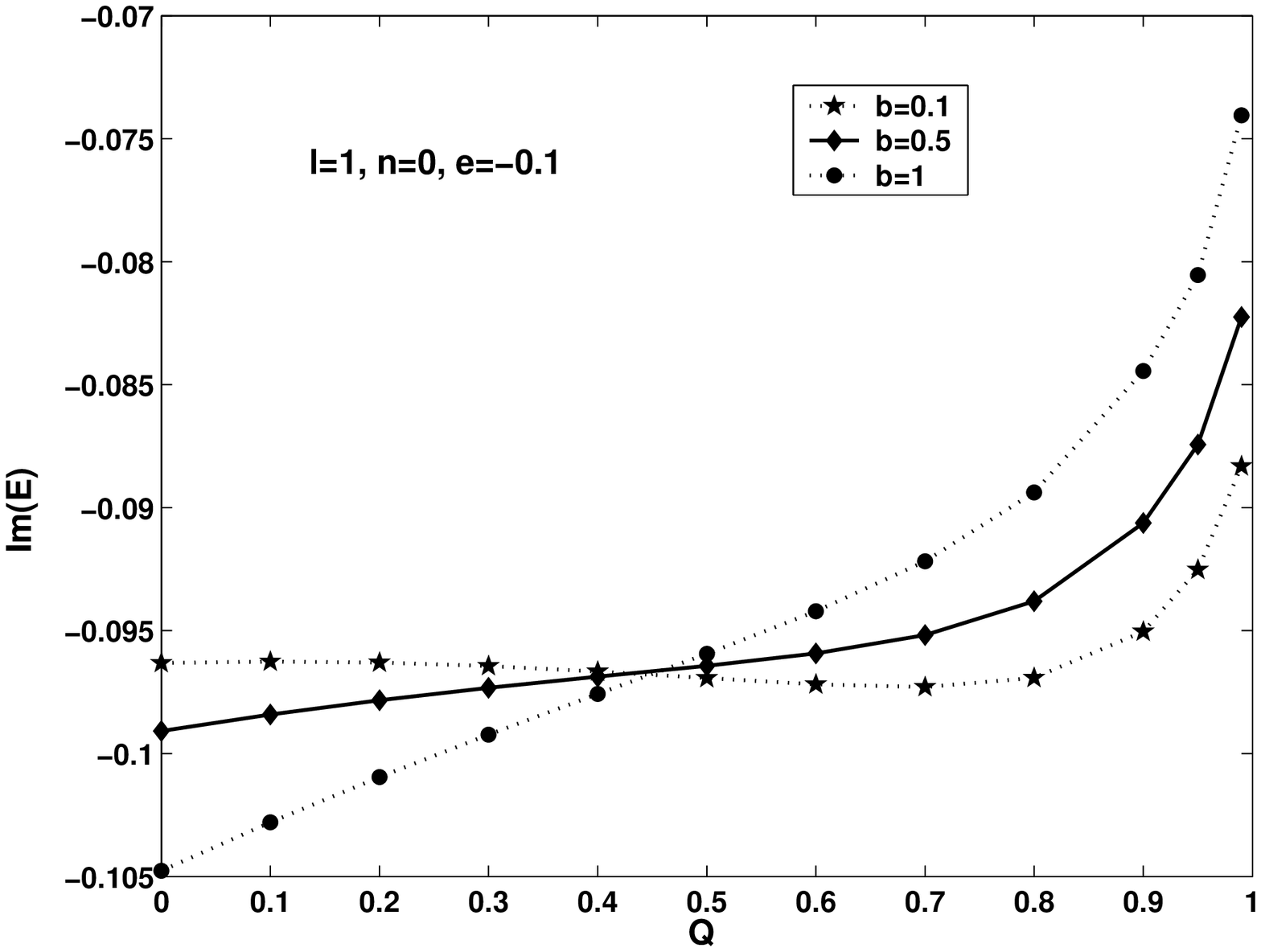}
\includegraphics[width=5.5cm]{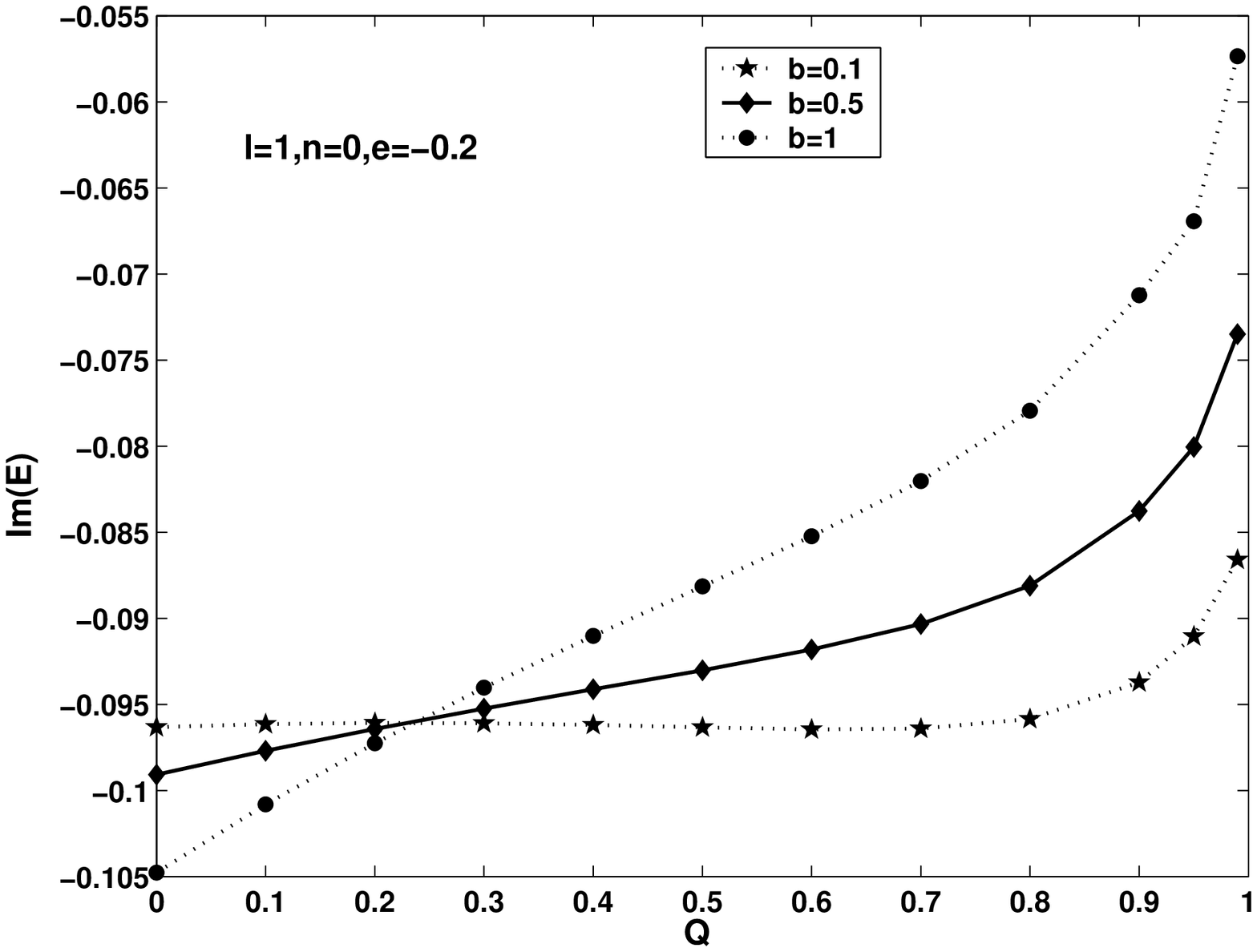}
\caption{Im(E) vs Q for l=1, n=0 for different e(+2 to -2)values
}\label{graph4}
\end{figure}

We now check for the imprint of cosmic string on the RN black hole.
For this we plot $Im(E)$ versus $Q$ graph for various $b$ values for
a fixed field charge $e$ as in Fig.\ref{graph4}. Thus for the mode
$l=1$, $n=0$ and when charge of the Dirac field is $e=0.2$, as $Q$
increases, $|Im(E)|$ at first increases and then decreases. But when
$b=0.1$ the $|Im(E)|$ is smaller compared to others. i.e., decay is
small in the case of black hole having cosmic string. This is true
only for the positively charged Dirac field and for $e=0$. When $e$
becomes negative, it shows another behavior. i.e., for the
negatively charged Dirac field, up to some $Q$ value (let it be
$Q_{0}$) $|Im(E)|$ for $b=0.1$ is small compared to others. The
$|Im(E)|$ for $b=0.1, 0.5$ and $1$ meets near $Q_{0}$. After that
$|Im(E)|$ values for $b=0.1$ become larger than all others. i.e., up
to $Q_{0}$ the decay rate is small for a black hole having cosmic
string and then decay rate increases.

\begin{figure}[h]\center
\includegraphics[width=5.5cm]{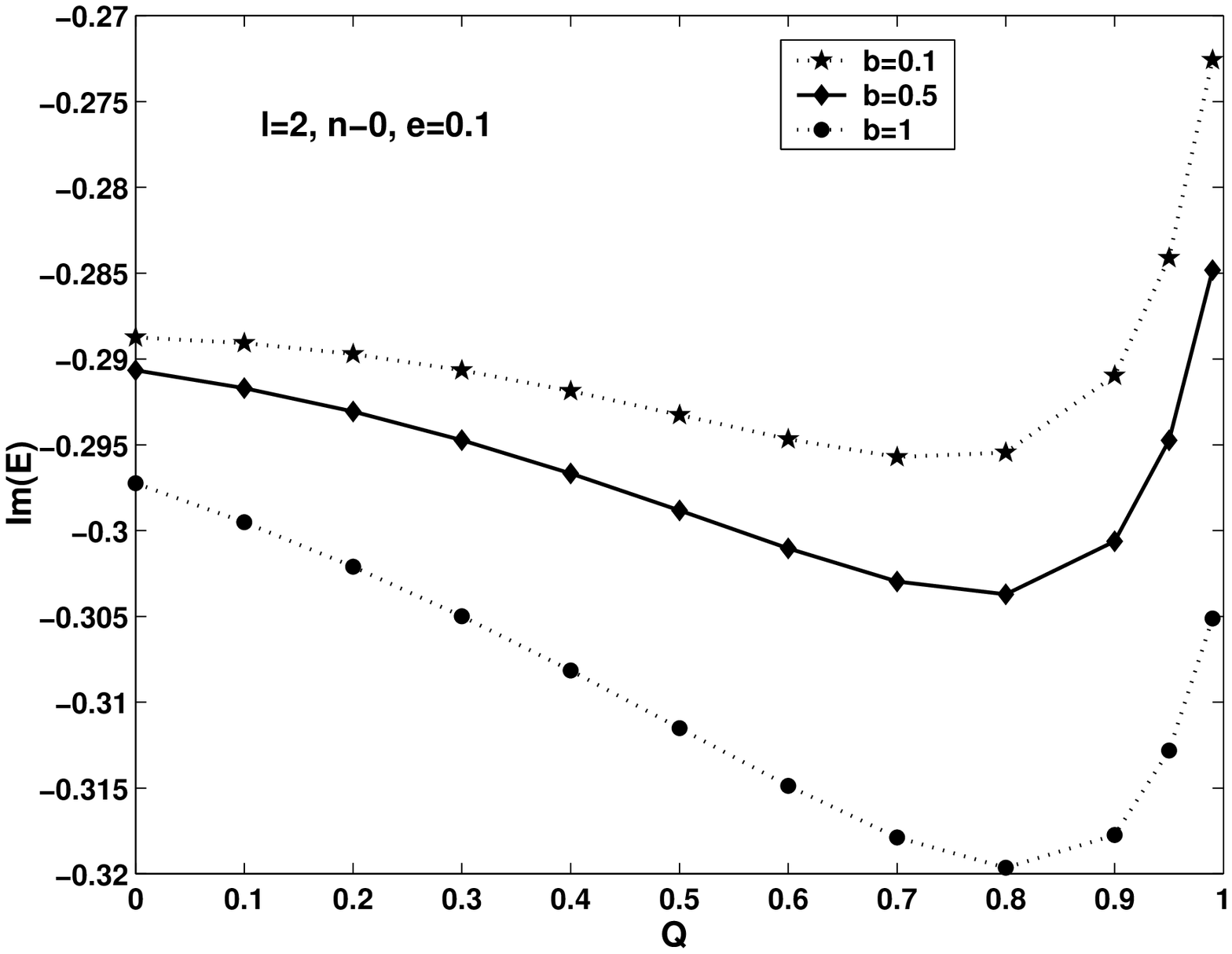}
\includegraphics[width=5.5cm]{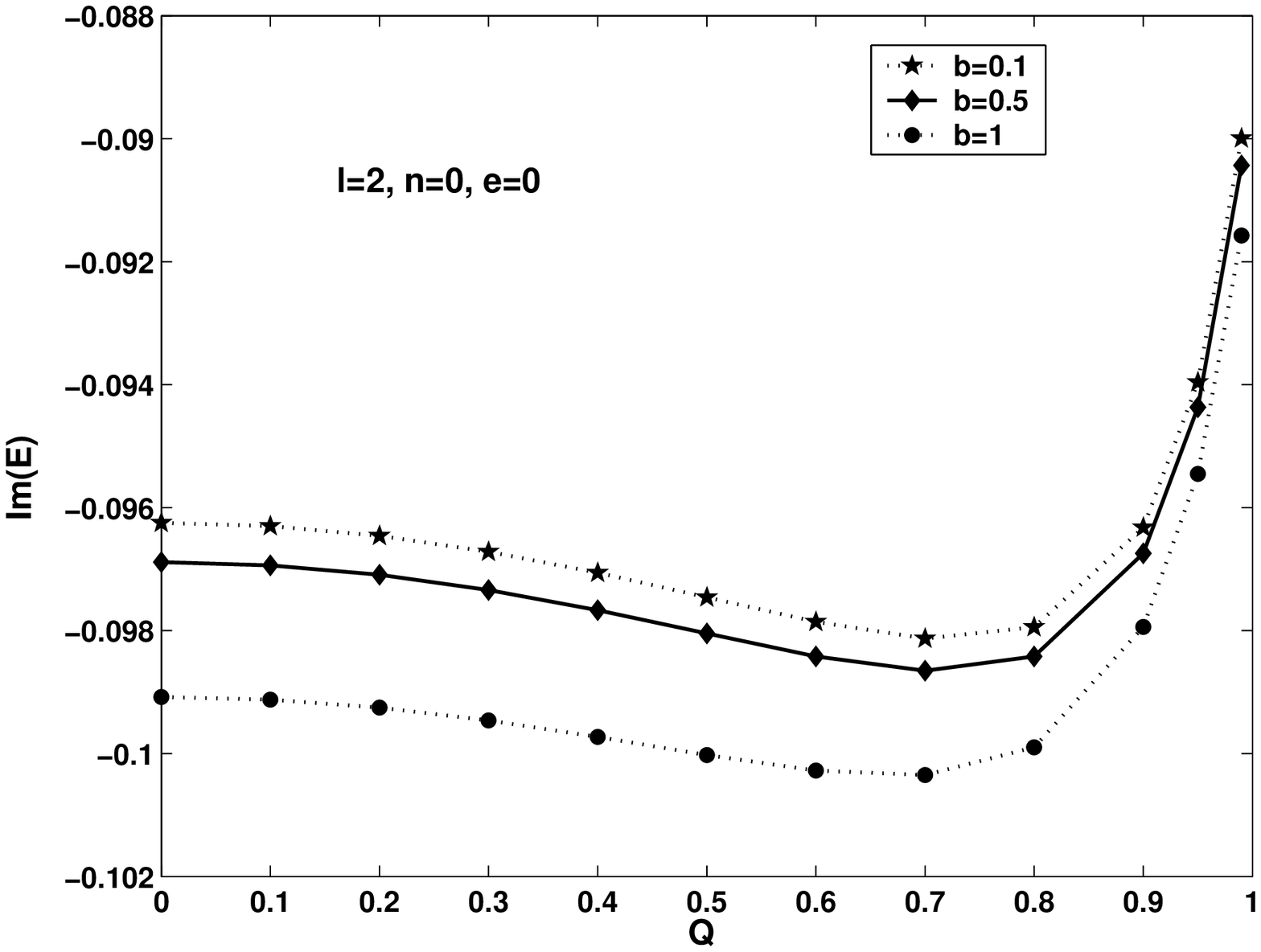}
\includegraphics[width=5.5cm]{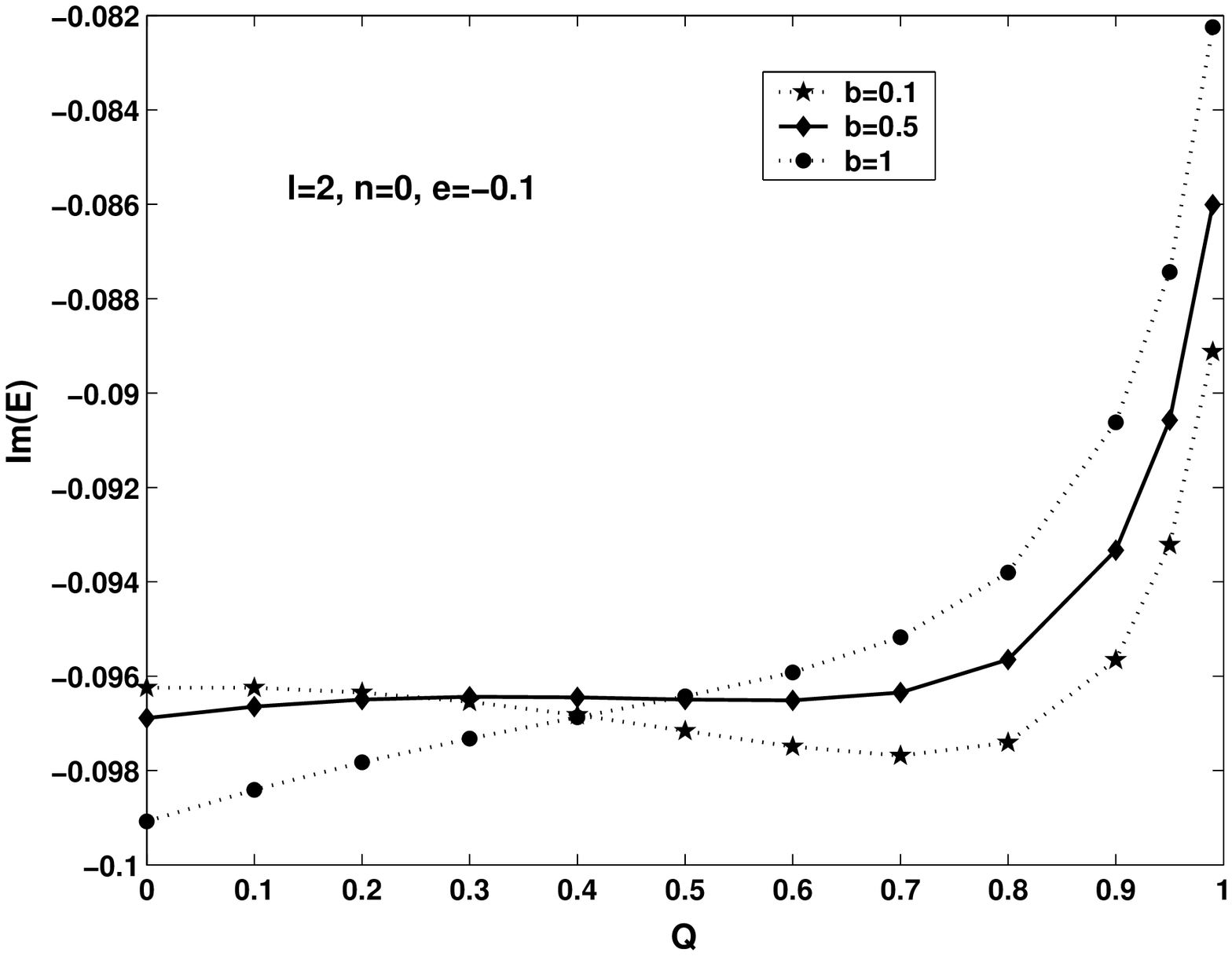}
\caption{Im(E) vs Q for l=2, n=0 for different e(+1 to
-1)values}\label{graph5}
\end{figure}

\begin{figure}[h]\center
\includegraphics[width=5.5cm]{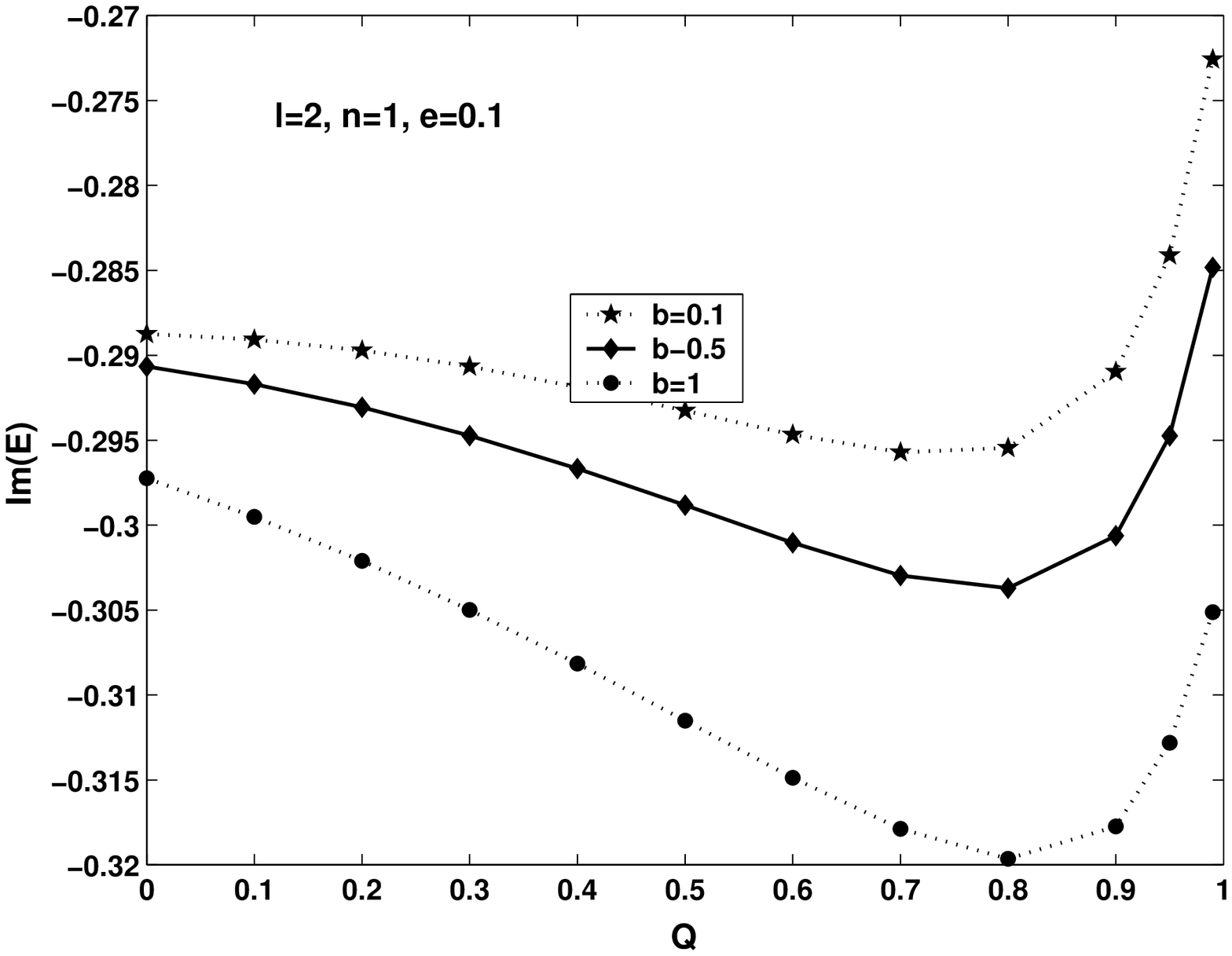}
\includegraphics[width=5.5cm]{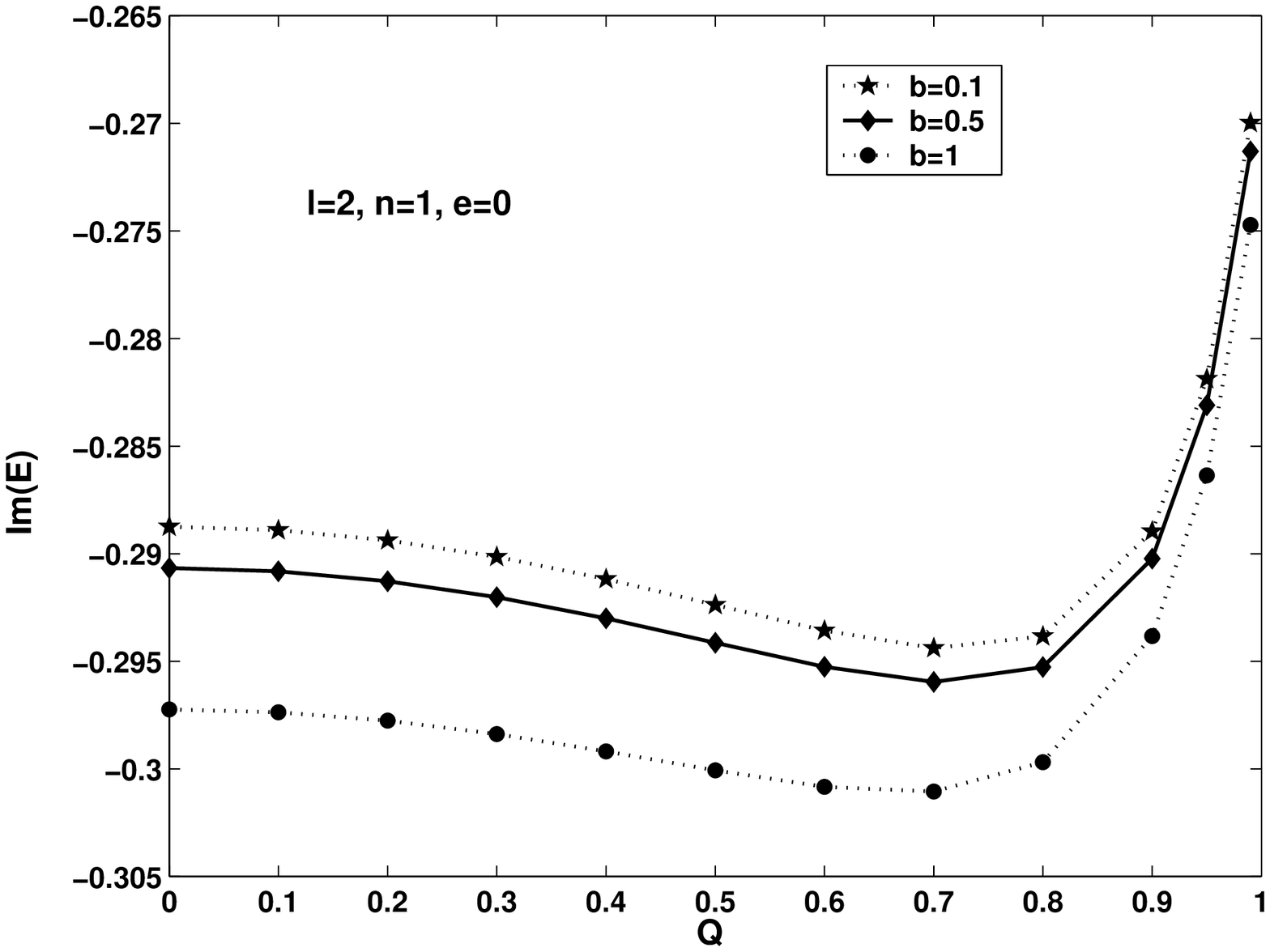}
\includegraphics[width=5.5cm]{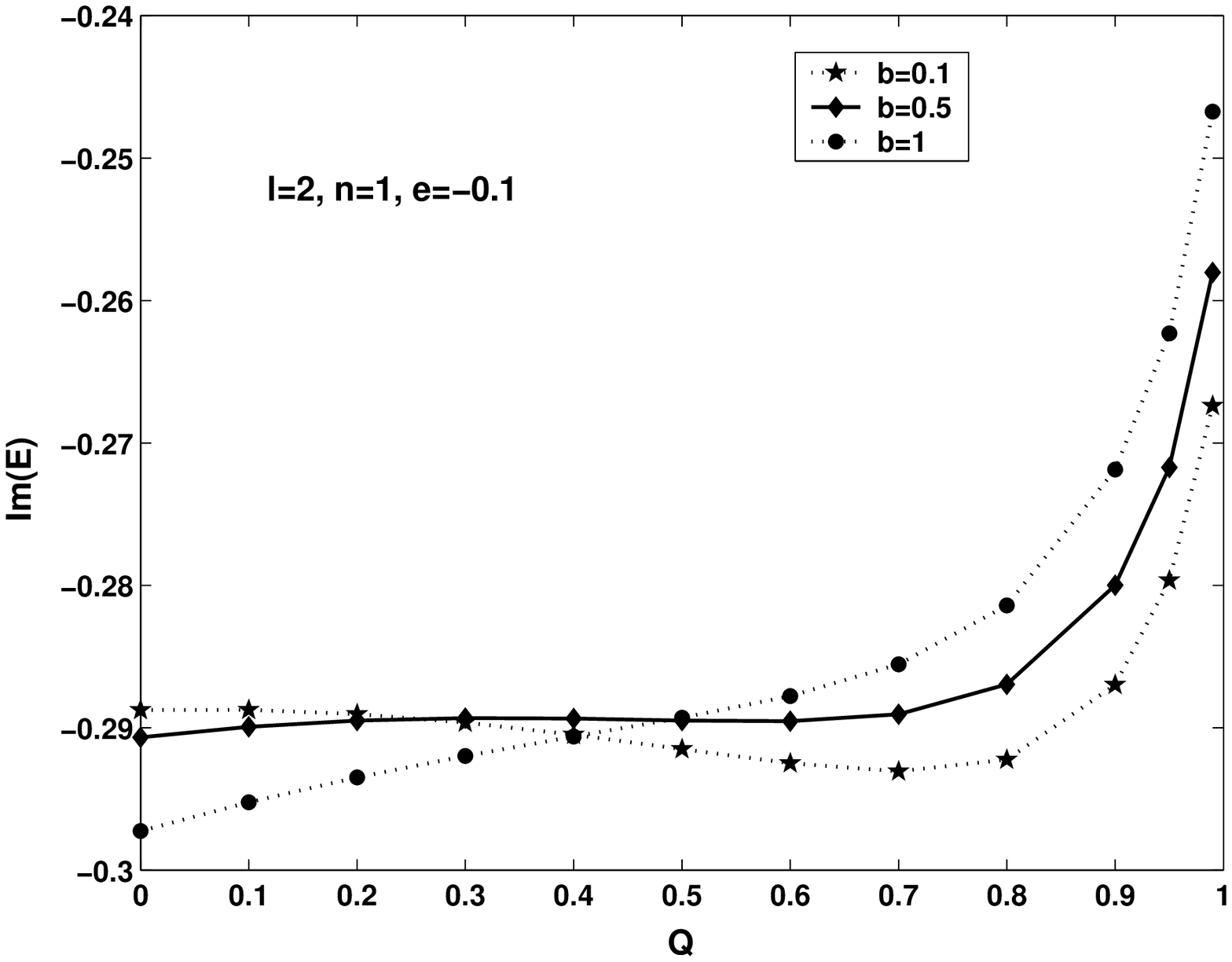}
\caption{Im(E) vs Q  for l=2, n=1 for different e(+1 to
-1)values}\label{graph6}
\end{figure}

The variation of $b$ with $e$ values for higher modes (for $l=2$,
$n=1$ and $n=2$) are shown in Fig. \ref{graph5} and Fig.
\ref{graph6}. Here also $|Im(E)|$ behaves same as above. i.e, in the
case of negatively charged Dirac field, when the charge of the black
hole is high, the effect due to cosmic string is suppressed.

\begin{figure}[h]\center
\includegraphics[width=5.5cm]{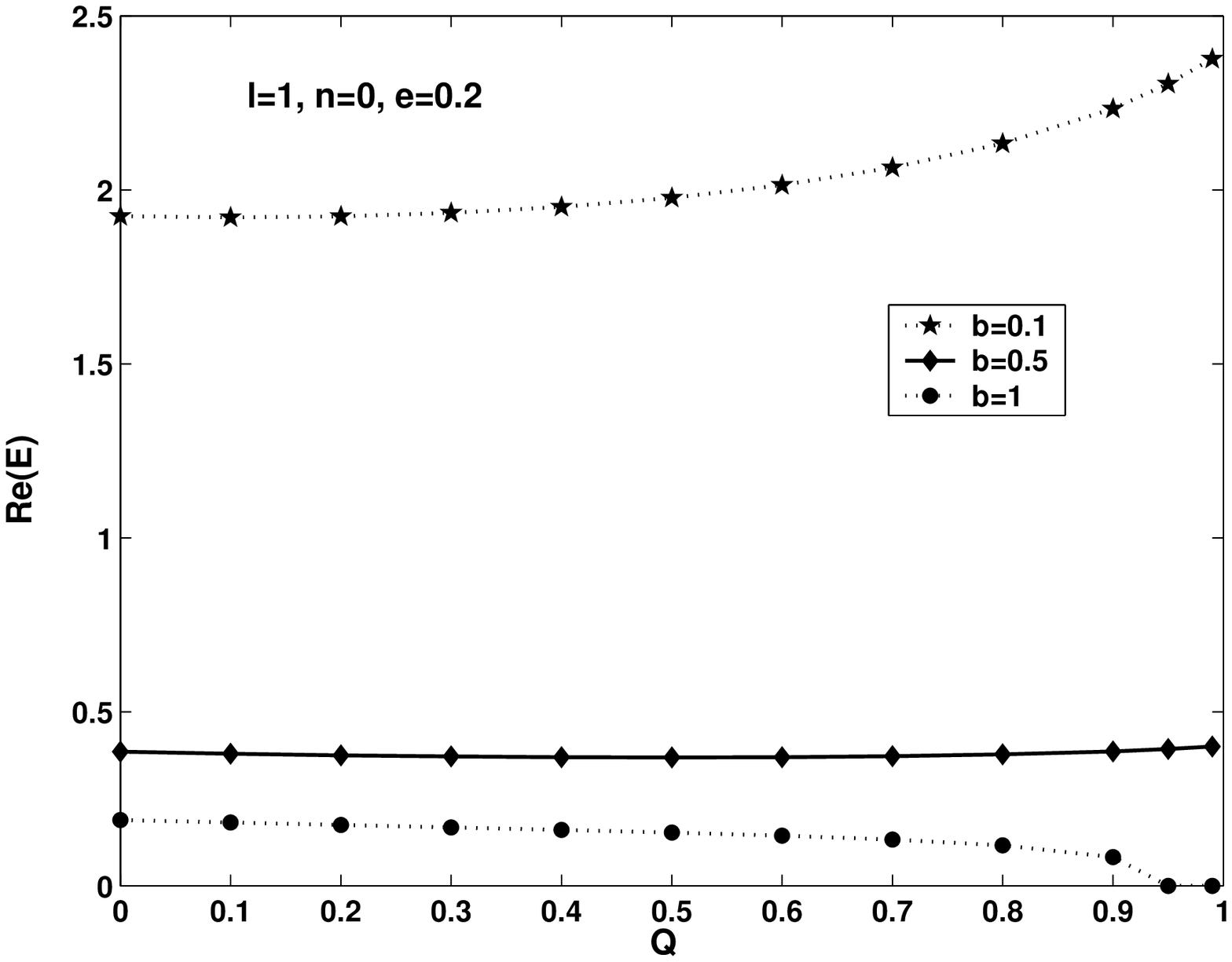}
\includegraphics[width=5.5cm]{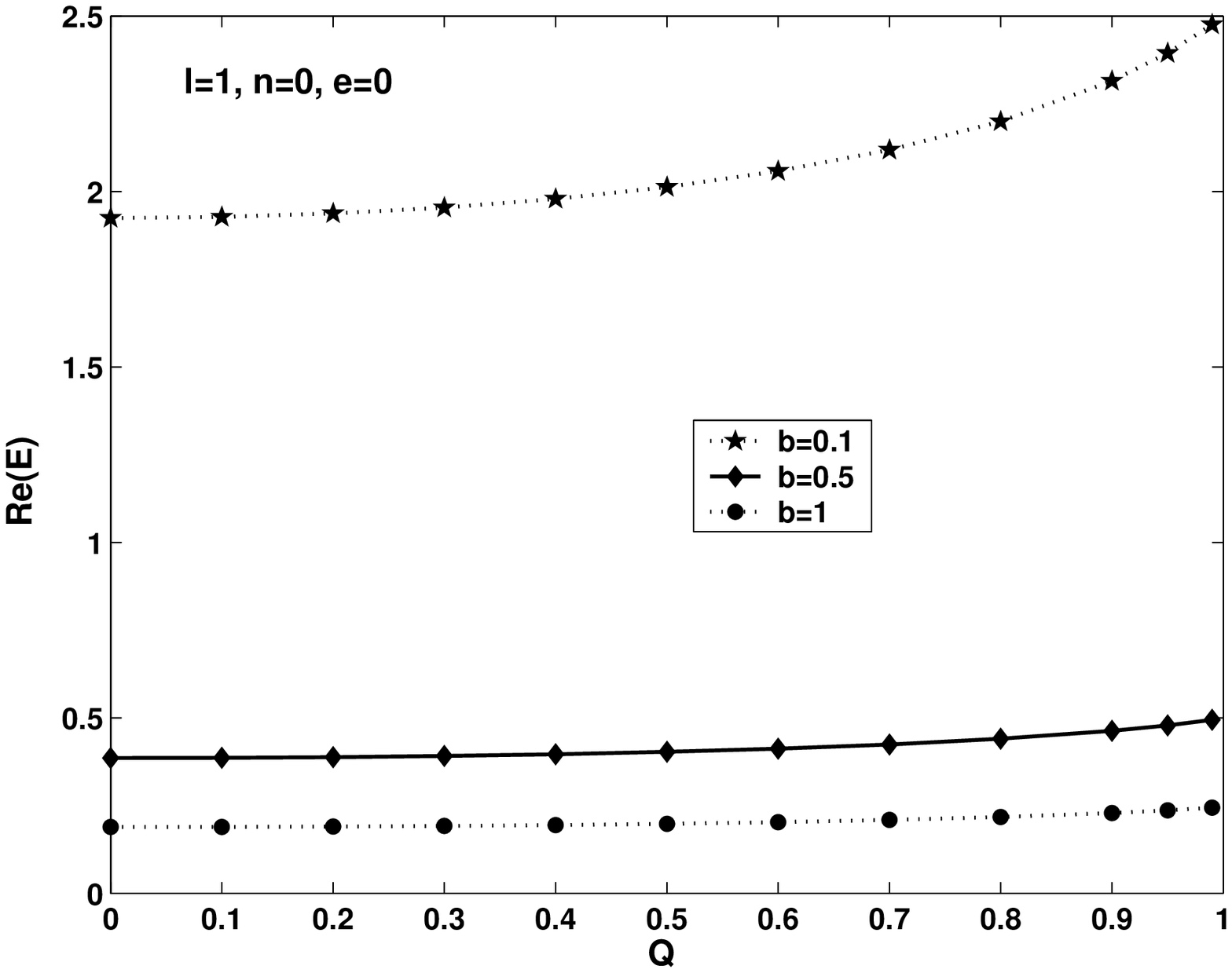}
\includegraphics[width=5.5cm]{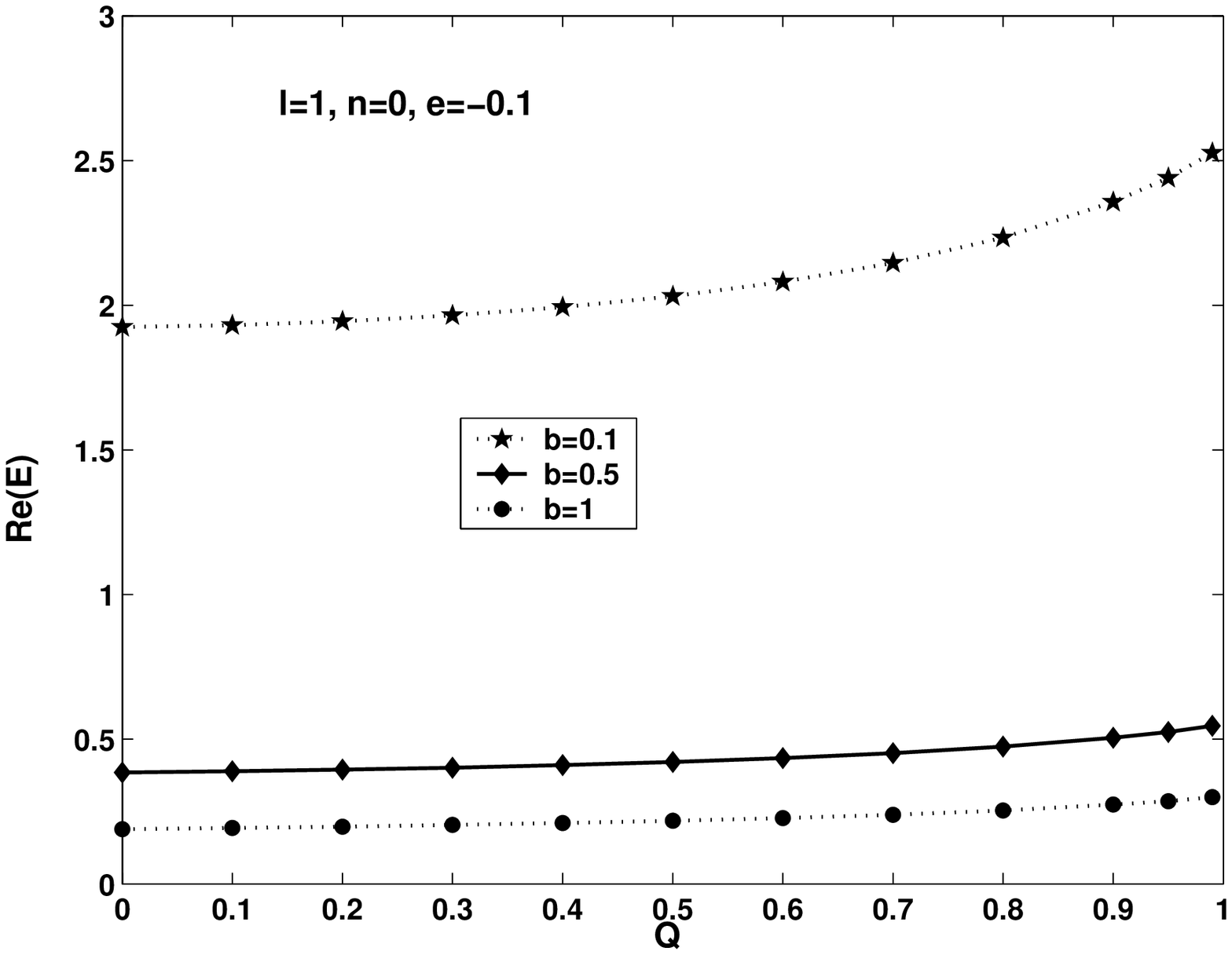}
\includegraphics[width=5.5cm]{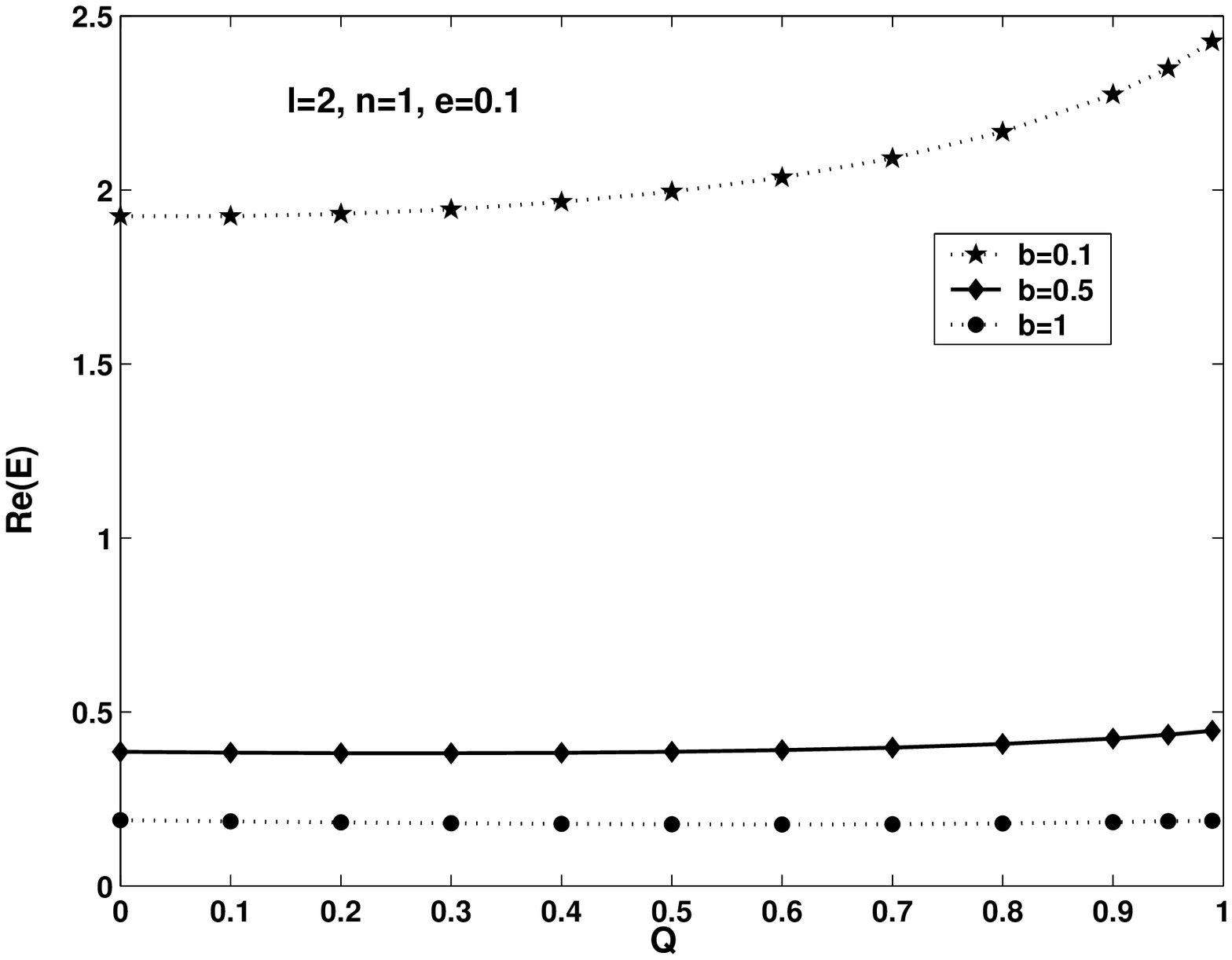}
\caption{Re(E) vs Q for e=+2, 0, -1 for l=1, and the last one is for
l=2}\label{graph7}
\end{figure}

Fig. \ref{graph7} shows the variation of $Re(E)$ with $Q$ for
various values of $b$ for a fixed $e$. Here for all values of $e$,
$Re(E)$ is larger for $b=0.1$ compared to others. That is, $Re(E)$
is larger for the black hole having cosmic string.

\section{Conclusion}
We have obtained the Dirac equation in RN black hole space time with
a cosmic sting and its deduction into a set of second order
differential equations. We have evaluated the quasinormal mode
frequencies for RN black hole space times having cosmic string
perturbed by a massless Dirac field. We find that for a fixed value
of $b$, positively charged Dirac field decay faster than negatively
charged Dirac field. But when we compare the RN black hole with and
without cosmic string, in the case of positively charged Dirac
field, decay is less when cosmic string is present. But in the case
of negatively charged Dirac field, the RN black hole having cosmic
string shows small decay for low values of black hole charge $Q$ but
as $Q$ increases its decay increases. Whereas, the RN black hole
which do not have cosmic string shows rapid decay for low values of
$Q$  and as $Q$ increases its decay rate decreases compared to the
RN black hole having cosmic string. Thus the effect due to cosmic
string will dominate only in the case of RN black hole having small
charge perturbed by a negatively charged Dirac field.

\section*{Acknowledgments}
The authors are thankful to U.G.C, New Delhi for financial support
through a Major Research Project. VCK wishes to acknowledge
Associateship of IUCAA, Pune. India.


\begin{thebibliography}{}

\bibitem{1f} T. Regge and J. A. Wheeler, Phys. Rev., 108, 1063 (1957).
\bibitem{1g} C. V. Vishveshwara , Phys. Rev. D, 1, 2870 (1970).
\bibitem{nlt99} H. P. Nollert, Class. Quantum. Grav., 16, R159 (1999).
\bibitem{kdk99} K. D. Kokkotas and Schmidt B. G., Living Rev. Rel., 2, 2 (1999).
\bibitem{1h} R. A. Konoplya, Phys. Lett. B, 550, 117 (2002).
\bibitem{bw05} B. Wang, Braz. J. Phys., 35, 1029 (2005).
\bibitem{1a} A. Vilenkin, Phys. Rep., 121, 263 (1986).
\bibitem{1b} T. W. B. Kibble, J. Phys. A: Math. Gen., 9, 1387 (1976).
\bibitem{1c} A. Vilenkin, Phys. Rev. Lett., 46, 1169 (1981); Phys. Rev. D, 24, 2082 (1981).
\bibitem{1d} T. W. B. Kibble and N. Turok ,  Phys. Rev. Lett., 116B, 141 (1982).
\bibitem{1e} A. Vilenkin,  Phys. Rev. D 23, 852 (1981).
\bibitem{sch07} S. Chen, B. Wang and R. Su, gr-qc/0701088v1 (2007).
\bibitem{sr08} R. Sini,  N. Varghese and V. C. Kuriakose, gr-qc/0802.0788v2 (2008).
\bibitem{ee} M. Aryal, L. H. Ford and A. Vilenkin, Phys. Rev. D, 34, 2263 (1986).
\bibitem{jnwheler} D. R. Brill and J. A. Wheeler Revs. Modern. Phys.,
29, 465 (1957)
\bibitem{1j} M.G. Germano, V. B. Bezerra and E. R. Bezerra de Mello, Class. Quantum Grav, 13, 2663 (1996).


\bibitem{aa} F. Cooper, A. Khare And U. Sukhatme, Phys. Rept., 251, 267 (1995).

\bibitem{vf84} V. Ferrari and B. Mashhoon , Phys. Rev. D, 30, 295
{1984}.
\bibitem{yuwu04}  Y. J. Wu and Z. Zhao, Phys. Rev. D, 69,
084015(2004).
\end{thebibliography}
\end{document}